\documentclass[10pt,journal,compsoc]{IEEEtran}
\usepackage{cite,mathtools}
\usepackage{graphicx}
\usepackage[tight,footnotesize]{subfigure}
\usepackage{amssymb}
\usepackage{url}
\usepackage{color}
\usepackage{epstopdf}

\ifCLASSOPTIONcompsoc
\else
\fi
\ifCLASSINFOpdf
\else
\fi
\begin{document}
%
\title{A Big Data Enabled Channel Model for 5G Wireless Communication Systems}
%
%
%
%

\author{Jie~Huang,
Cheng-Xiang~Wang,~\IEEEmembership{Fellow,~IEEE},
Lu~Bai,
Jian~Sun,~\IEEEmembership{Member,~IEEE},
Yang~Yang,~\IEEEmembership{Fellow,~IEEE},
Jie~Li,~\IEEEmembership{Senior~Member,~IEEE},
Olav~Tirkkonen,~\IEEEmembership{Member,~IEEE},
and Ming-Tuo~Zhou,~\IEEEmembership{Senior~Member,~IEEE}
\IEEEcompsocitemizethanks{\IEEEcompsocthanksitem J. Huang and C.-X. Wang are with National Mobile Communications Research Laboratory, School of Information Science and Engineering, Southeast University, Nanjing 210096, China, and also with
the Purple Mountain Laboratories, Nanjing 211111, China. E-mail: j\_huang@seu.edu.cn, chxwang@seu.edu.cn. (Corresponding authors: Yang Yang and Cheng-Xiang Wang)

\IEEEcompsocthanksitem L. Bai is with School of Cyber Science and Technology, Beihang University, Beijing 100191, China. E-mail: lu\_bai@buaa.edu.cn.

\IEEEcompsocthanksitem J. Sun is with Shandong Provincial Key Lab of Wireless Communication Technologies, School of Information Science and Engineering, Shandong University, Qingdao 266237, China. E-mail: sunjian@sdu.edu.cn.

\IEEEcompsocthanksitem Y. Yang is with Shanghai Institute of Fog Computing Technology (SHIFT), ShanghaiTech University, Shanghai 201210, China. E-mail: yangyang@shanghaitech.edu.cn.

\IEEEcompsocthanksitem J. Li is with Department of Computer Science and Engineering, Shanghai Jiaotong University, Shanghai 200240, China. E-mail: lijiecs@sjtu.edu.cn.

\IEEEcompsocthanksitem O. Tirkkonen is with Aalto University, Espoo 02150, Finland. E-mail: olav.tirkkonen@aalto.fi.

\IEEEcompsocthanksitem M.-T. Zhou is with Shanghai Institute of Microsystem and Information Technology (SIMIT), Chinese Academy of Sciences (CAS), Shanghai 200050, China. E-mail: mingtuo.zhou@mail.sim.ac.cn.
}
}

%
%

\markboth{IEEE Transactions on big data,~Vol.~x, No.~x, Month~2020}%
{Shell \MakeLowercase{\textit{et al.}}: Bare Demo of IEEEtran.cls for Computer Society Journals}
%



\IEEEtitleabstractindextext{%
\begin{abstract}
The standardization process of the fifth generation (5G) wireless communications has recently been accelerated and the first commercial 5G services would be provided as early as in 2018. The increasing of enormous smartphones, new complex scenarios, large frequency bands, massive antenna elements, and dense small cells will generate big datasets and bring 5G communications to the era of big data. This paper investigates various applications of big data analytics, especially machine learning algorithms in wireless communications and channel modeling. We propose a big data and machine learning enabled wireless channel model framework. The proposed channel model is based on artificial neural networks (ANNs), including feed-forward neural network (FNN) and radial basis function neural network (RBF-NN). The input parameters are transmitter (Tx) and receiver (Rx) coordinates, Tx--Rx distance, and carrier frequency, while the output parameters are channel statistical properties, including the received power, root mean square (RMS) delay spread (DS), and RMS angle spreads (ASs). Datasets used to train and test the ANNs are collected from both real channel measurements and a geometry based stochastic model (GBSM). Simulation results show good performance and indicate that machine learning algorithms can be powerful analytical tools for future measurement-based wireless channel modeling.
\end{abstract}

}

\maketitle

\IEEEdisplaynontitleabstractindextext

%
\IEEEpeerreviewmaketitle

\IEEEraisesectionheading{\section{Introduction}\label{sec:introduction}}

%
%
%
%
\IEEEPARstart{W}{ith} the rapid increasing of smartphones and versatile new applications, the mobile data grows exponentially in recent years. The Cisco visual networking index (VNI) released the white paper about global mobile data traffic forecast for 2016--2021 in February 2017 \cite{Cisco17}. In summary, global mobile data traffic grew 18-fold over the last 5 years. It grew 63\% in 2016 and reached 7.2 exabytes (EBs) per month at the end of 2016. Mobile devices and connections grew to 8.0 billion in 2016. The annual mobile data traffic will exceed half a zettabyte (ZB) by 2021. The main trends contributing to the growth of mobile data traffic include evolving towards smarter mobile devices and advanced cellular networks.

In recent years, the wireless communication network has dramatical improvements to support the huge mobile data traffic. The fifth generation (5G) wireless communication network is expected to greatly improve the data rate by 1000 times, reduce the latency, and achieve higher energy and cost efficiencies \cite{Wang14, Yang17}. The standardization process of 5G systems has recently been accelerated and the first commercial 5G services would be provided as early as in 2018 \cite{And14}. 5G will be applied in enhanced mobile broadband (eMBB), massive machine type communication (mMTC), and ultra reliable and low latency communications (uRLLC) scenarios \cite{ITU}. In order to achieve this goal, 5G will be a paradigm shift that includes very high carrier frequencies with large bandwidths, unprecedented numbers of antennas, and extreme base station and device densities \cite{And14, Shafi17}. Millimeter wave (mmWave), massive multiple-input multiple-output (MIMO), and ultra-dense networks (UDNs) have been seen as ``big three'' potential key technologies to achieve the goal of the 5G wireless communication systems \cite{Ge16}. The increasing of enormous smartphones, new complex scenarios, large frequency bands, massive antenna elements, and dense small cells will generate big datasets and bring 5G wireless communications to the era of big data \cite{Bi15,Fer16}.

The term ``big data'' became popular and widespread as recently as in 2011. Volume, variety, and velocity have been a common framework to describe features of big data in the early stage \cite{Gan15}. Other dimensions like veracity and value have been added to describe big data later \cite{Katal13}, leading to the popular five V's. 

Big data analytics have usually been applied to areas such as text, image, audio, video, social media, and predictive analytics \cite{Gan15}. For a 5G mobile network, as the data grows rapidly, it will bring lots of challenges and opportunities when acquiring, storing, and processing the wireless big data \cite{Chin14}. Compared with the big datasets in traditional areas, wireless big data has some additional properties, and big data analytics may not be applied to wireless communications directly. Apart from the aforementioned five V's, wireless big data is distinct in its unique multi-dimensional, personalized, multi-sensory, and real-time features \cite{Cheng17, Cheng17_2}. The multi-dimensional spatio-temporal data contains the information of user trajectories. Also, the mobile data is highly personalized and relevant to the user's location and context, and is usually obtained from multiple sensors in real-time. Wireless data traffic contains strong correlative and statistical features in various dimensions including time, location, and the underlying social relationship \cite{Bi15}. 

Discovering the relationship between big data analytics and wireless communications has been a challenging task \cite{Imran14, Han15, Zheng16, Jiang16, Han17, Qian17, Ahm18, Zhang18}. Some attempts have been made to apply big data analytics to the area of wireless communications. Authors in \cite{Qian17} separated existing researches in wireless big data into data, transmission, network, and application layers. Wireless channel modeling, which is the foundation of wireless communications, is related to transmission layer. When signals are transmitted by the transmitter (Tx), they will undergo serious distortions. The faded signals are then received by the receiver (Rx) through direct transmission, reflection, scattering, and diffraction. The signals are characterized by a number of multipath components (MPCs) with parameters of complex amplitude, delay, Doppler shift, and departure and arrival angles. The MPC parameters are highly correlated with the network layout, including Tx and Rx locations, carrier frequency, and scatterer distributions, etc. Thus, important channel statistical properties such as the received power, root mean square (RMS) delay spread (DS), and RMS angle spread (AS) may have a complicated non-linear relationship with the network layout. 

Inspired by the good learning and prediction performance of artificial neural network (ANN) which has been throughly investigated, we propose an ANN based channel model framework. Both the feed-forward neural network (FNN) and radial basis function neural network (RBF-NN) are used to predict important channel statistical properties. Compared with existing deterministic and stochastic channel modeling approaches, the proposed channel model has a good trade-off among accuracy, complexity, and flexibility. Existing channel models rely on many assumptions, while ANN based channel model framework is directly learned from the datasets and can be more accurate. For different network layout configurations (carrier frequency, Tx/Rx position, etc.), the existing channel models should be run each time, which is complicated and time consuming. On contrary, the channel statistical properties can be directly obtained in a simple way by the learned machine/function in real-time. Moreover, different environments should be constructed in ray tracing model and different parameter sets should be obtained in WINNER-like model each time, while a more general ANN based channel model framework can be learned from datasets collected from various scenarios. The performance of the proposed ANN based channel model is fully investigated through extensive simulations based on real channel measurement data and geometry based stochastic model (GBSM) generated data.

The remainder of this paper is organized as follows. Section~II surveys different machine learning algorithms and shows an overview of big data analytics in wireless communications and channel modeling. In Section~III, some basic knowledge about ANN is given, and the ANN based channel model framework is proposed.  Simulation results based on real channel measurement data and GBSM generated data are then analyzed in Section IV, which validates the ANN based channel model framework, and extensions and discussions of the proposed channel model are also given. Finally, conclusions are drawn and some future research directions are given in Section V.

\section{Overview of Big Data Analytics in Wireless Communications}
\label{sec_BigAnalytics}

Generally, big data analytical tools include stochastic modeling, data mining, and machine learning \cite{Bi15}. Stochastic modeling uses probabilistic models to capture the explicit features and dynamics of the data traffic. Data mining focuses on exploiting the implicit structures in the mobile dataset. Machine learning can establish a functional relationship between input data and output actions, thus achieving auto-processing capability for unseen patterns of data inputs \cite{Bi15}. Specifically, machine learning algorithms have developed dramatically over the past few years and have been applied to various areas. We concentrate on investigating the applications of machine learning algorithms in wireless communications and channel modeling in this paper.

\subsection{Different Machine Learning Algorithms}

In general, machine learning algorithms can be simply categorized as supervised learning, unsupervised learning, and reinforcement learning. Supervised learning aims to learn the mapping from the input data to the output data. For unsupervised learning, the goal is to model the underlying structure of the input data. For reinforcement learning, simple reward feedback is required to automatically determine the ideal behavior within a specific context to maximize its performance. Some types of popular machine learning algorithms include decision tree, Bayesian, clustering, classification, regression, dimensionality reduction, ANN, deep learning algorithms, etc. The detailed descriptions and representatives of these algorithms are shown in Table \ref{tab:algo}. Note that classification and regression are two main purposes of machine learning. Other types of algorithms listed in Table \ref{tab:algo} can also be applied to classification and regression problems. The main difference between classification and regression is that the output variable takes class labels for classification problems, but it takes continuous values instead for regression problems.

\begin{table*}[tb!]
\caption{Different types of machine learning algorithms.}
\label{tab:algo}
\begin{center}
\begin{tabular}{|c|p{5cm}|p{6.6cm}|}
\hline
Machine learning algorithms&Descriptions&Representatives\\
\hline
Decision tree&Use a tree-like graph or model to learn decision rules&Iterative Dichotomiser 3 (ID3), C4.5, classification and regression tree (CART)\\
\hline
Bayesian& Use the Bayes rule to infer model parameters&Naive Bayesian, Bayesian network\\
\hline
Clustering&Group similar data points into a cluster&K-means, K-nearest neighbors (KNN), fuzzy C-means (FCM), DBSCAN, expectation maximization (EM)\\
\hline
Classification&Classify data into given set of categories& AdaBoost, support vector machine (SVM), relevance vector machine (RVM)\\
\hline
Regression&Learn the relationship between variables in the dataset&LASOO, minimax probability machine (MPM), Gaussian process regression (GPR)\\
\hline
Dimensionality reduction&Exploit the inherent structure of the dataset to describe it with less information&Principal component analysis (PCA), random forests\\
\hline
ANN&A processing network with amounts of neurons inspired by neuroscience&FNN, RBF-NN, multilayer perceptron (MLP), extreme learning machine (ELM), wavelet neural network (WNN)\\
\hline
Deep learning&Use a cascade of many layers of non-linear processing units for feature extraction and transformation& Convolution neural network (CNN), recurrent neural network (RNN), deep belief network (DBN), deep Boltzmann machine (DBM)\\
\hline
\end{tabular}
\end{center}
\end{table*}

\subsection{Machine Learning in Wireless Communications}
Recently, there have been various developments in applying big data analytics to wireless communications. The core to use machine learning algorithms in wireless communications is that many problems in wireless communications can be converted to clustering, classification, and regression problems. By learning and training the big datasets in wireless communications, the network can be more intelligent to achieve better performance and adapt to various applications. 

A popular application of machine learning in wireless communications is indoor and outdoor localization/positioning \cite{Mar10, Ngu15, Zou16, Liang17, Ye17}. In \cite{Mar10}, features that represent propagation conditions were extracted from ultra-wideband (UWB) measurement data, then classification and regression algorithms were developed based on SVM. SVM can be used for non-probabilistic binary classification. The developed algorithms were capable of non-line-of-sight (NLOS) identification and mitigation to reduce localization errors. In \cite{Ngu15}, the RVM based localization algorithm was applied to NLOS identification and mitigation. Compared to SVM, RVM has an identical functional form but provides probabilistic classification based on Bayesian statistics. In \cite{Zou16, Liang17}, two ELM based algorithms were developed for indoor positioning. ELM is a FNN for classification or regression with a single layer of hidden nodes, where the weights connecting inputs to hidden nodes are randomly assigned and never updated. The weights between hidden nodes and outputs are learned in a single step, which essentially amounts to learning a linear model. In \cite{Ye17}, a FNN based localization algorithm was proposed using channel fingerprint vectors as inputs. In \cite{Thi13}, the SVM was used for spectrum sensing. In \cite{Joung16}, the SVM was used for antenna selection, while in \cite{He18}, both the SVM and Naive Bayesian were utilized for antenna selection. Other applications include caching \cite{Bas14, Bas15, Yao16}, resource allocation \cite{Yang16, Huang17, Bao17}, interference management \cite{Deb15}, channel estimation \cite{He18_2, Tang18, Luo18}, modulation classification \cite{Meng18}, scenario classification \cite{Alh18}, user clustering \cite{Cui18}, etc.

\subsection{Machine Learning in Wireless Channel Modeling}

There have been a new trend to apply big data analytics to wireless channel modeling \cite{Chin14}. As new complex scenarios, large frequency bands, and massive antennas will be used for 5G wireless communication systems, big channel impulse response (CIR) datasets will be generated from channel measurement campaigns. Meanwhile, many new channel characteristics should be measured and modeled, including three-dimensional (3-D) double-directional angles, non-stationarity in spatial-temporal-frequency domains, spherical wavefront, high path loss, and high delay resolution.

We summarize recent developments and applications of machine learning algorithms in wireless channel modeling, as shown in Table \ref{tab:mac_ch}. In \cite{Chang97}, the RBF-NN was used to predict the path loss. The Tx and Rx heights, Tx--Rx distance, carrier frequency, and intercede range were the inputs, and the output was the path loss. In \cite{Liu12}, the sparse Bayesian learning of RVM was applied to direction of arrival (DoA) estimation. It first obtained coarse signal locations with the sparsity-inducing RVM on a predefined spatial grid, and then achieved refined direction estimation via searching. In \cite{Braz13}, the RVM was employed to filter the MPCs of measured power delay profiles (PDPs) in indoor environments, enabling the determination of the delays and complex amplitudes. The RVM used few kernel functions to generate the sparsity concept, and it allowed the estimation of channel parameters as well as the number of MPCs. In \cite{Azp14}, the FNN and RBF-NN were combined with ray launching in complex indoor environments. The neural network was used to predict the intermediate points in ray launching algorithm to decrease the computation complexity. In \cite{Zhang16}, big data was used to model wireless channels, and a cluster-nuclei based channel model was proposed. First, the measurement data was processed by using high resolution estimation algorithms to obtain MPC parameters and then clustered. Meanwhile, the image processing algorithms were applied to reconstruct the measurement environment and find main scattering objects. The clusters and scatterers were matched based on the cluster characteristics and object properties, and then a limited number of cluster-nuclei were formed. With the cluster-nuclei, the CIR was produced by machine learning algorithms such as ANN. With the big datasets, the CIR prediction in various scenarios can be realized. In \cite{Ma17,Ma17_2}, the ANN was used to remove the noise from measured CIR, and the PCA was utilized to exploit the features and structures of the channel and model the CIR. In \cite{Fer16_2}, the MLP was applied to predict the received signal strength. The inputs were Tx--Rx distance and diffraction loss, while the output was the  received signal strength. In \cite{Li17}, the CNN was used to automatically identify different wireless channels and help decide which relevant wireless channel features should be used. The MPC parameters like amplitude, delay, and Doppler frequency were extracted using the space-alternating generalized expectation-maximization (SAGE) algorithm and used as input parameters in the CNN, and the output of the CNN was the class of the wireless channels. In \cite{Bla17}, the complex LASSO algorithm was applied to estimate the tapped delay line CIR from channel measurement data. In \cite{Ucc18}, the SVM was used to predict the path loss for smart metering applications. It was based on received signal strength measurements and a 3-D map of the propagation environment. In \cite{He18_3}, various clustering algorithms were used for clustering and tracking of MPCs, including K-means, FCM, and DBSCAN.

\begin{table}[tb!]
\caption{Applications of machine learning in wireless channel modeling.}
\label{tab:mac_ch}
\begin{center}
\begin{tabular}{|p{0.6cm}|p{1.3cm}|p{5.5cm}|}
\hline
Ref.&Algorithms&Applications\\
\hline
\cite{Chang97} &RBF-NN & Predict the path loss\\
\hline
\cite{Liu12} &RVM & Estimate DoA of MPCs\\
\hline
\cite{Braz13}&RVM&Filter the noise embedded MPCs to determine the PDP\\
\hline
\cite{Azp14}& FNN and RBF-NN&Predict the intermediate points in ray launching simulation to obtain the indoor received power\\
\hline
\cite{Zhang16}&ANN&Produce the CIR with a limited number of cluster-nuclei\\
\hline
\cite{Ma17,Ma17_2}&ANN and PCA&Remove the noise and estimate the CIR\\
\hline
\cite{Fer16_2} &MLP & Predict the outdoor received signal strength \\
\hline
\cite{Li17}&CNN&Extract channel features and identify different wireless channels based on channel measurement data\\
\hline
\cite{Bla17}&LASSO&Estimate the CIR based on channel measurement data\\
\hline
\cite{Ucc18}&SVM&Predict the path loss\\
\hline
\cite{He18_3}&K-means, FCM, and DBSCAN&Clustering and tracking MPCs\\
\hline
\end{tabular}
\end{center}
\end{table}

\section{A Machine Learning Enabled Channel Model}
Channel modeling is important for system design and performance evaluation. From channel modeling, some important channel statistical properties including large-scale and small-scale parameters can be obtained. Generally, channel measurements will be indispensable to validate channel models. For the coming 5G wireless communications, the scenarios become more complicated, such as mmWave, massive MIMO, high-speed trains, etc. Channel measurements should be conducted to study new channel propagation characteristics in these challenging scenarios. The huge bandwidths, massive antennas, fast velocity, and various scenarios will generate big datasets which are time consuming for data post-processing and needed to be handled by machine learning. By learning from channel measurement datasets, important channel statistical properties can be obtained and expressed as a non-linear function of arbitrary known inputs, thus decreasing the time consuming channel measurements and complicated data post-processing works. However, a channel sounder which is able to satisfy all the 5G new deployments is very expensive and challenging, and channel measurement campaigns are also very time consuming. It is hard to achieve big datasets from channel measurement campaigns, which should contain various configurations such as different scenarios, Tx and Rx antenna coordinates, Tx--Rx distances, and carrier frequencies. We resort to both real channel measurement datasets and GBSM simulation datasets. Measurement datasets are obtained at some fixed locations, while simulation datasets are obtained in a random manner.

\subsection{Channel Measurement Datasets}
\begin{figure}[tb!]
\centering
\includegraphics[width=3.3in]{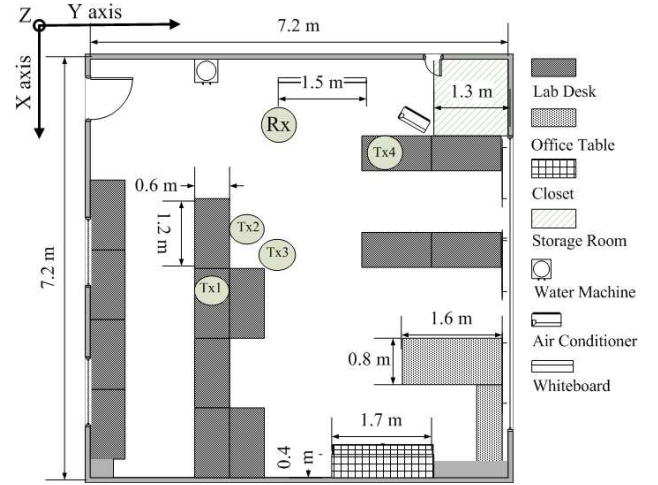}
\caption{Layout and antenna locations of the measurement environment.}
\label{fig:layout}
\end{figure}

Channel measurement campaigns were conducted in an indoor office environment with room size of 7.2$\times$7.2$\times$3~$\rm{m}^3$, as shown in Fig. \ref{fig:layout} \cite{JSAC17}. Four mmWave frequency bands were measured, i.e., 11, 16, 28, and 38 GHz bands. In each band, the Rx antenna was located at (1,3,1.45) and scanned in a large uniform rectangular array (URA), and Tx antennas were located at Tx1(4,2.2,2.6), Tx2(3.2,2.4,2.6), Tx3(3.6,3,2.6), and Tx4(2,5.2,2.6). Antenna elements in the URA for the four bands are 51$\times$51, 76$\times$76, 91$\times$91, and 121$\times$121, respectively. The sweeping points for 11 GHz and 16 GHz bands is 401, while it is 801 for 28 GHz and 38 GHz bands. The SAGE algorithm is used to extract MPC parameters from the sub-array measurement data. The sub-array is 10$\times$10, 15$\times$15, 15$\times$15, and 20$\times$20 for the four bands, respectively. More details about the channel measurements and data processing can be found in \cite{JSAC17}.

\subsection{GBSM Simulation Datasets}
GBSM is a popular channel modeling approach. It represents scatterers with regular or irregular geometry shapes. The MPC parameters including power, delay, and departure and arrival angles are then calculated from the geometry relationships.

The GBSM used here is a 3-D wideband mmWave MIMO channel model \cite{SCIS18}. Clusters in the 3-D space are generated with a homogeneous Poisson point process (PPP) to capture effects of the environment. The Tx and Rx antennas are surrounded by two spheres to mimic the clustering property of MPCs in both delay and angular domains. The scatterers are related with rays in each cluster which are confined by surfaces of a circular cone and a spherical segment. After the generation of all clusters and scatterers, routes from the Tx antenna to the Rx antenna are formed, each corresponds to a MPC with parameters of received power, delay, and departure and arrival angles. The received power is calculated from path loss model, the delay is obtained from the length of each MPC, while the departure and arrival angles are calculated from the geometry relationships. The channel model is validated by comparing with channel measurements in the same office environment and used to generate big channel datasets.

Denote the Tx coordinate as $(x_t,y_t,z_t)$, the Rx coordinate as $(x_r,y_r,z_r)$, the Tx--Rx distance as $d_{tr}$, and the carrier frequency as $f_c$. The eight parameters are known and used as a represent of the environment.

A total of $L$ MPCs are simulated from the GBSM with parameters [$p_l$,$\tau_l$,$\phi_{tl}$,$\theta_{tl}$,$\phi_{rl}$,$\theta_{rl}$], $l=1,...,L$, which are the received power, delay, azimuth angle of departure (AAoD), elevation angle of departure (EAoD), azimuth angle of arrival (AAoA), and elevation angle of arrival (EAoA) of the $l$-th MPC, respectively. From the MPC parameters, some important channel statistical properties can be obtained. The total received power is calculated as
\begin{equation}
P=\sum_{l=1}^{L}p_l.
\end{equation}

The RMS DS is a important second-order statistic to describe channel dispersion in delay domain and can be calculated as
\begin{equation}
DS=\sqrt{\frac{\sum_{1}^{L}p_l\tau _l^{2}}{\sum_{1}^{L}p_l}-(\frac{\sum_{1}^{L}p_l\tau _l}{\sum_{1}^{L}p_l})^{2}}.
\end{equation}

The RMS AS is a important second-order statistic to describe channel dispersion in angle domain and can be calculated as
\begin{equation}
AS=\sqrt{\frac{\sum_{1}^{L}p_l\psi _l^{2}}{\sum_{1}^{L}p_l}-(\frac{\sum_{1}^{L}p_l\psi _l}{\sum_{1}^{L} p_l})^{2}}
\end{equation}
where $\psi _l$ denotes either of the four angles.

\subsection{ANN Based Channel Model}
We share the similar viewpoint with research works in the literature and use ANN for wireless channel modeling. ANN is based on a large collection of simple neural units. Each neural unit is connected with many others. Thresholds of the nodes and weights of the connections are trained to learn the non-linear relationship between the inputs and outputs. The datasets are usually divided as training and testing datasets.

For multi-dimensional modeling, FNN and RBF-NN are the extensively used ANN algorithms. FNN increases the number of non-linear segmentations using two layers containing activation functions to achieve high non-linearity. FNN contains one input layer, one hidden layer, and one output layer. RBF-NN uses a single layer with a very high number of neurons with the same goal of achieving higher number of non-linear segmentation. Because neurons are added as learning continues and only one layer of weights is to be adjusted for RBF-NN, its computational cost of learning is less than that for FNN \cite{Azp14}.

The channel measurement data and GBSM simulation data are both used to train and test the neural network. For channel measurement data, there are 25, 25, 36, and 36 sub-arrays for the four bands, respectively. In addition, four Tx positions were measured, thus obtaining 100 (25$\times$4), 100 (25$\times$4), 144 (36$\times$4), and 144 (36$\times$4) groups of datasets for the four bands, respectively. The measurement datasets are divided into 400 groups of training datasets and 88 testing datasets. For GBSM simulation data, a total of 500 groups of datasets are generated from 500 times Monte-Carlo simulations, in which 400 groups are used for training and the left 100 groups are used for testing. The Tx and Rx coordinates are varying randomly in the confined environment, the Tx--Rx distance is calculated according to Tx and Rx coordinates, and the carrier frequency is varying randomly in the range of 10 GHz to 60 GHz. The eight parameters are used as inputs of the ANN. The outputs are the six channel statistical properties, including received power $P$, RMS DS $\sigma_{ds}$, AAoD spread (ADS) $\sigma_{ads}$, EAoD spread (EDS) $\sigma_{eds}$, AAoA spread (AAS) $\sigma_{aas}$, and EAoA spread (EAS) $\sigma_{eas}$. The relationship between the inputs and outputs is
\begin{equation}
y(P,\sigma _{ds},\sigma _{ads},\sigma _{eds},\sigma _{aas},\sigma _{eas})=f(x_t,y_t,z_t,x_r,y_r,z_r,d_{tr},f_c).
\end{equation}

As the inputs have definite physical meanings, their values are varying in different ranges and at different levels with different units, the input parameters should be normalized and mapped to be in the range of $-$1 to 1. The outputs also have similar inverse conversion operations to obtain real predicted values.

\begin{figure}[tb!]
\centering
\includegraphics[width=3in]{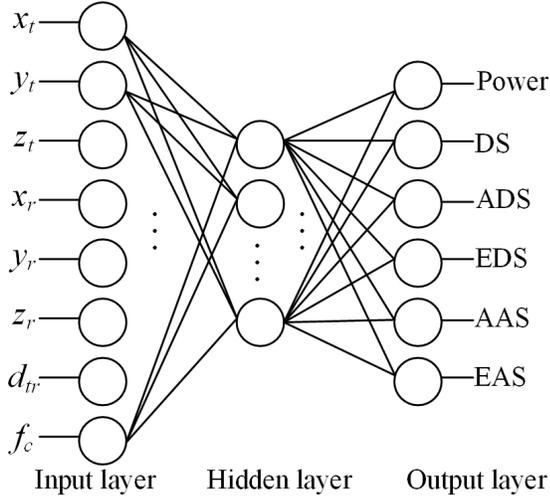}
\caption{The general ANN based channel model framework.}
\label{fig:ann}
\end{figure}

The general ANN based channel model framework is illustrated in Fig. \ref{fig:ann}. In our simulations, both the FNN and RBF-NN are used to predict channel statistical properties. The prediction performance of the two ANN algorithms are also compared.

\section{Results and Analysis}

The MATLAB neural network toolbox is used to build the neural network. The FNN and RBF-NN are constructed by `newff' and `newrb' functions, respectively.

For FNN, the `logsig', `tansig', and `purelin' activation functions are used for input layer, hidden layer, and output layer, respectively. The activation functions are given as

\begin{equation}
logsig(n)=\frac{1}{1+e^{-n}}
\end{equation}

\begin{equation}
tansig(n) = \frac{2}{1+e^{-2n}}-1
\end{equation}

\begin{equation}
purelin(n)=n.
\end{equation}

The number of neurons in the hidden layer of FNN is selected from trial and error. Different number of neurons in the hidden layer are tested and 10 hidden neurons is found to achieve good performance with relatively low complexity. Thus, a 8-10-6 FNN is constructed. The goal of convergence error is set as $5\times10^{-2}$. The weight vectors are trained to achieve the goal of convergence error.

For RBF-NN, the goal of convergence error is set as $3\times10^{-2}$. The Gaussian RBF is given as
\begin{equation}
G(n)=e^{-\frac{(n-c)^2}{2\sigma^2}}
\end{equation}
where $c$ is the center point, and $\sigma$ is the spread of Gaussian RBF which is set as 0.75 to have a smooth interpolation plane. In the training procedure, 5 neurons are added between two displays.

In the simulations, for both FNN and RBF-NN, the Levenberg-Marquardt optimization algorithm is used to train the neural network, the mean square error (MSE) is selected to evaluate the prediction performance, and the number of iterations was set as 1000. Compared with RBF-NN, the goal of convergence for FNN is larger to avoid local optimization and over-fitting. The detailed parameters and values for FNN and RBF-NN are summarized in Table \ref{tab:para_va}.

\begin{table}[btb!]
\caption{Parameters and values for FNN and RBF-NN.}
\label{tab:para_va}
\begin{center}
\begin{tabular}{|p{3cm}|p{5cm}|}
\hline
Parameters&Values\\
\hline
Optimization algorithm &Levenberg-Marquardt algorithm\\
\hline
Performance metric &Mean square error (MSE)\\
\hline
Number of iterations&1000\\
\hline
The goal of convergence error& FNN: $5\times10^{-2}$ (to avoid local optimization and over-fitting); RNF-NN: $3\times10^{-2}$\\
\hline
Number of neurons in the hidden layer&FNN: 10 (tuned in practice); RBF-NN: 5 neurons are added between two displays when training\\
\hline
Spread of Gaussian RBF&0.75 (to have a smooth interpolation plane)\\
\hline
\end{tabular}
\end{center}
\end{table}

Once the ANN is trained to achieve the goal of convergence error, the neural network is well learned for the training datasets, and it can predict the outputs for testing input parameters. The predicted outputs can then be compared with measured and simulated outputs to evaluate the performance of the learned neural network.

\subsection{Predicting Statistical Properties for Measurement Datasets}
\begin{figure}[tb!]
\centering
\includegraphics[width=3.4in]{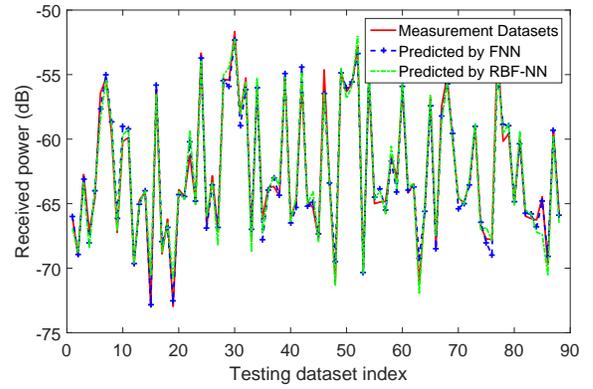}
\caption{Measurement and predicted received power.}
\label{fig:mea_pl}
\end{figure}

\begin{figure}[tb!]
\centering
\includegraphics[width=3.4in]{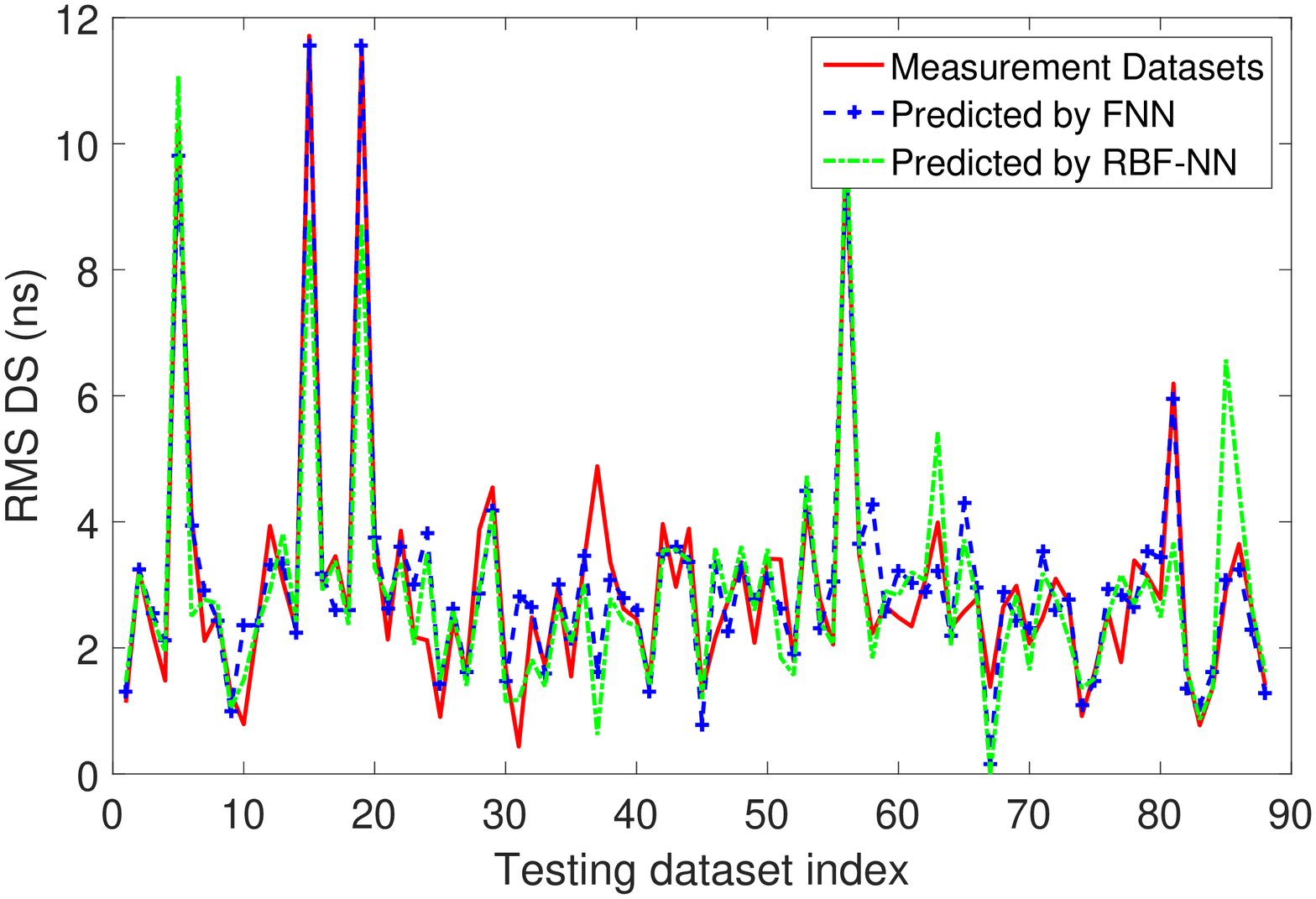}
\caption{Measurement and predicted RMS DS.}
\label{fig:mea_ds}
\end{figure}

\begin{figure}[tb!]
\centering
\includegraphics[width=3.4in]{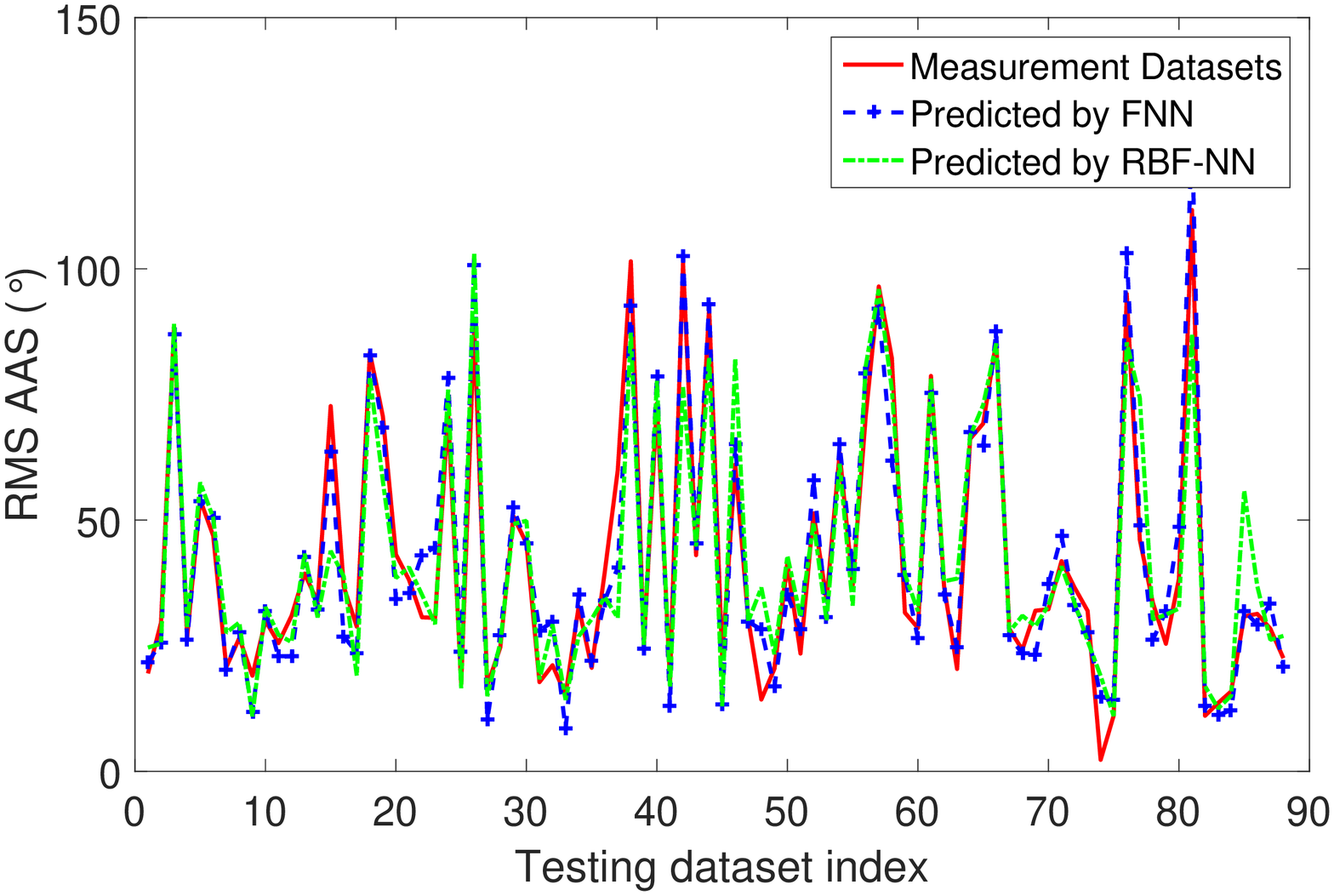}
\caption{Measurement and predicted RMS AAS.}
\label{fig:mea_aas}
\end{figure}

\begin{figure}[tb!]
\centering
\includegraphics[width=3.4in]{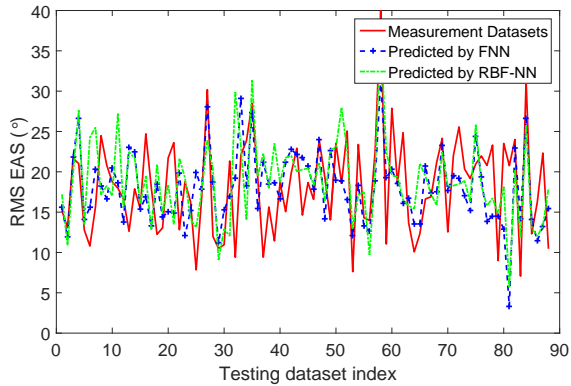}
\caption{Measurement and predicted RMS EAS.}
\label{fig:mea_eas}
\end{figure}

Important channel statistical properties including received power, RMS DS, AAS, and EAS are predicted. In Fig. \ref{fig:mea_pl}, the measurement and predicted received powers are shown. Both the prediction performances of FNN and RBF-NN are illustrated. The x-axis denotes the testing dataset index. The measurement and predicted RMS DSs are illustrated in Fig.~\ref{fig:mea_ds}. The measurement and predicted RMS AASs and EASs are illustrated in Fig. \ref{fig:mea_aas} and Fig. \ref{fig:mea_eas}, respectively. From these results, we can see that both the FNN and RBF-NN work well on the measurement datasets. The received power, RMS DS, and RMS AAS are accurately predicted, while the prediction error for RMS EAS is slightly larger.

\subsection{Predicting Statistical Properties for Simulation Datasets}

\begin{figure}[tb!]
\centering
\includegraphics[width=3.4in]{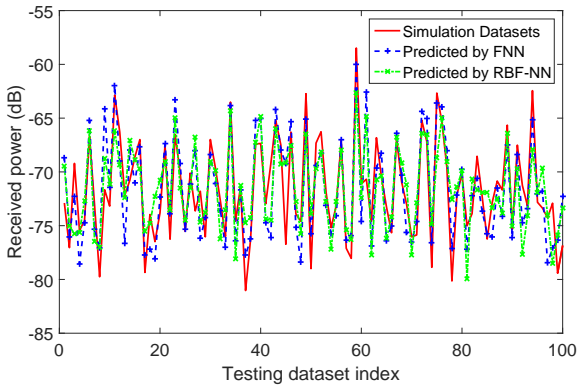}
\caption{Simulated and predicted received power.}
\label{fig:sim_pl}
\end{figure}

\begin{figure}[tb!]
\centering
\includegraphics[width=3.4in]{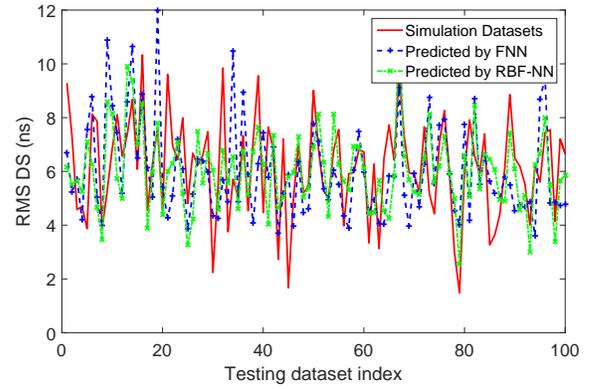}
\caption{Simulated and predicted RMS DS.}
\label{fig:sim_ds}
\end{figure}

\begin{figure*}[t!]
\centering
\begin{minipage}[t]{0.48\linewidth}
\centerline{\includegraphics[width=3in]{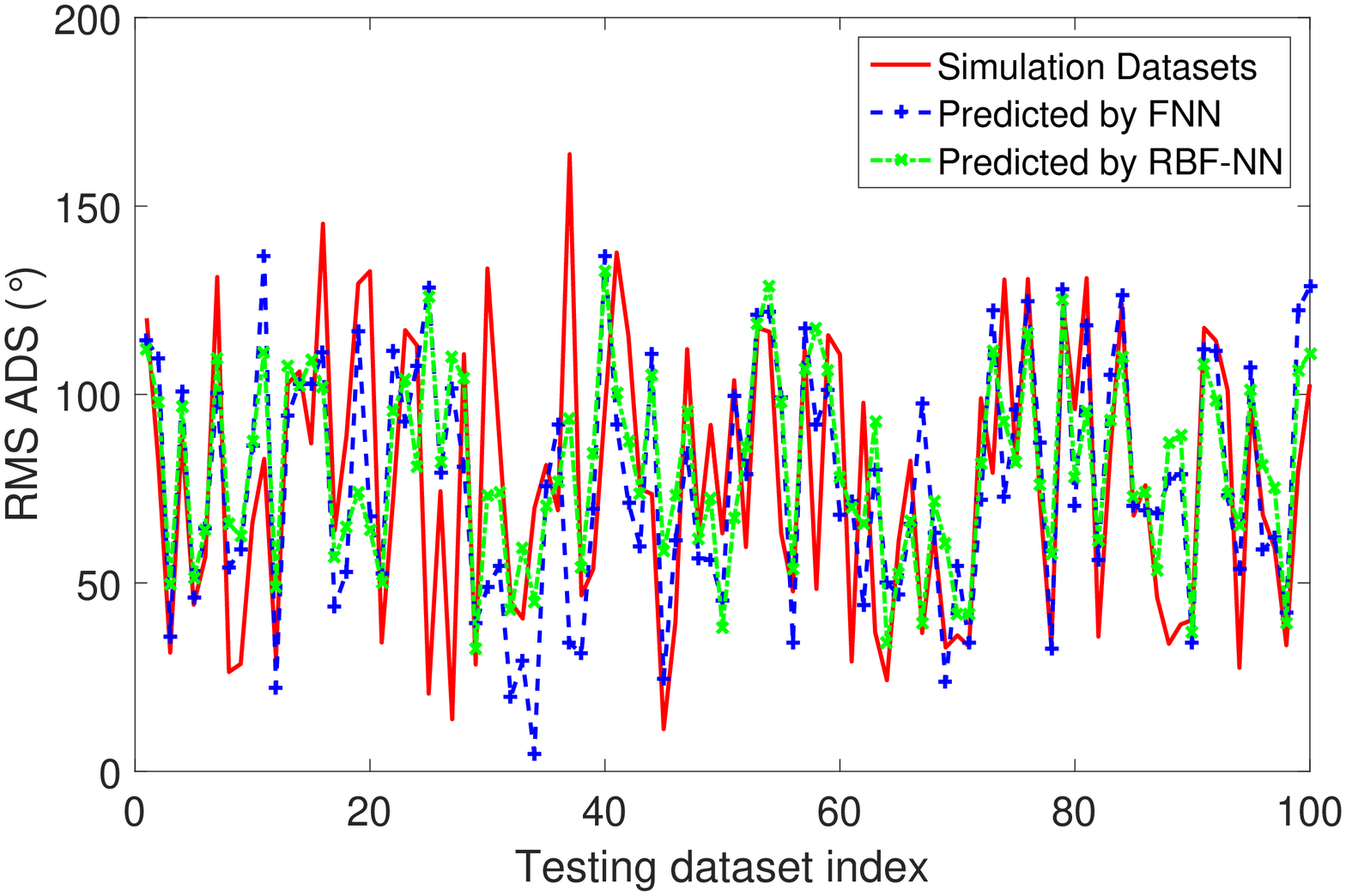}}
\footnotesize \centerline{(a)}
\end{minipage}
\begin{minipage}[t]{0.48\linewidth}
\centerline{\includegraphics[width=3in]{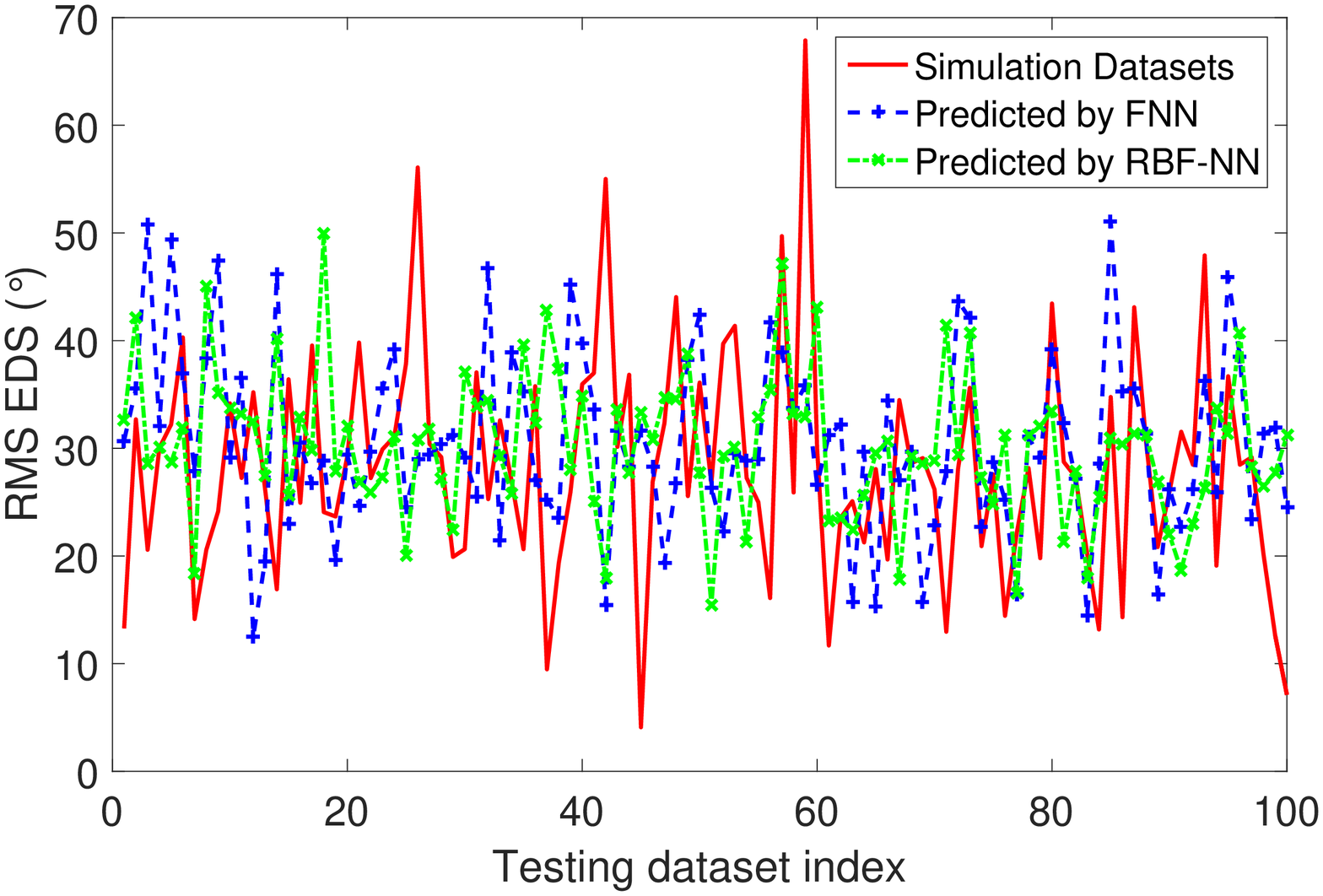}}
\footnotesize \centerline{(b)}
\end{minipage}

\begin{minipage}[t]{0.48\linewidth}
\centerline{\includegraphics[width=3in]{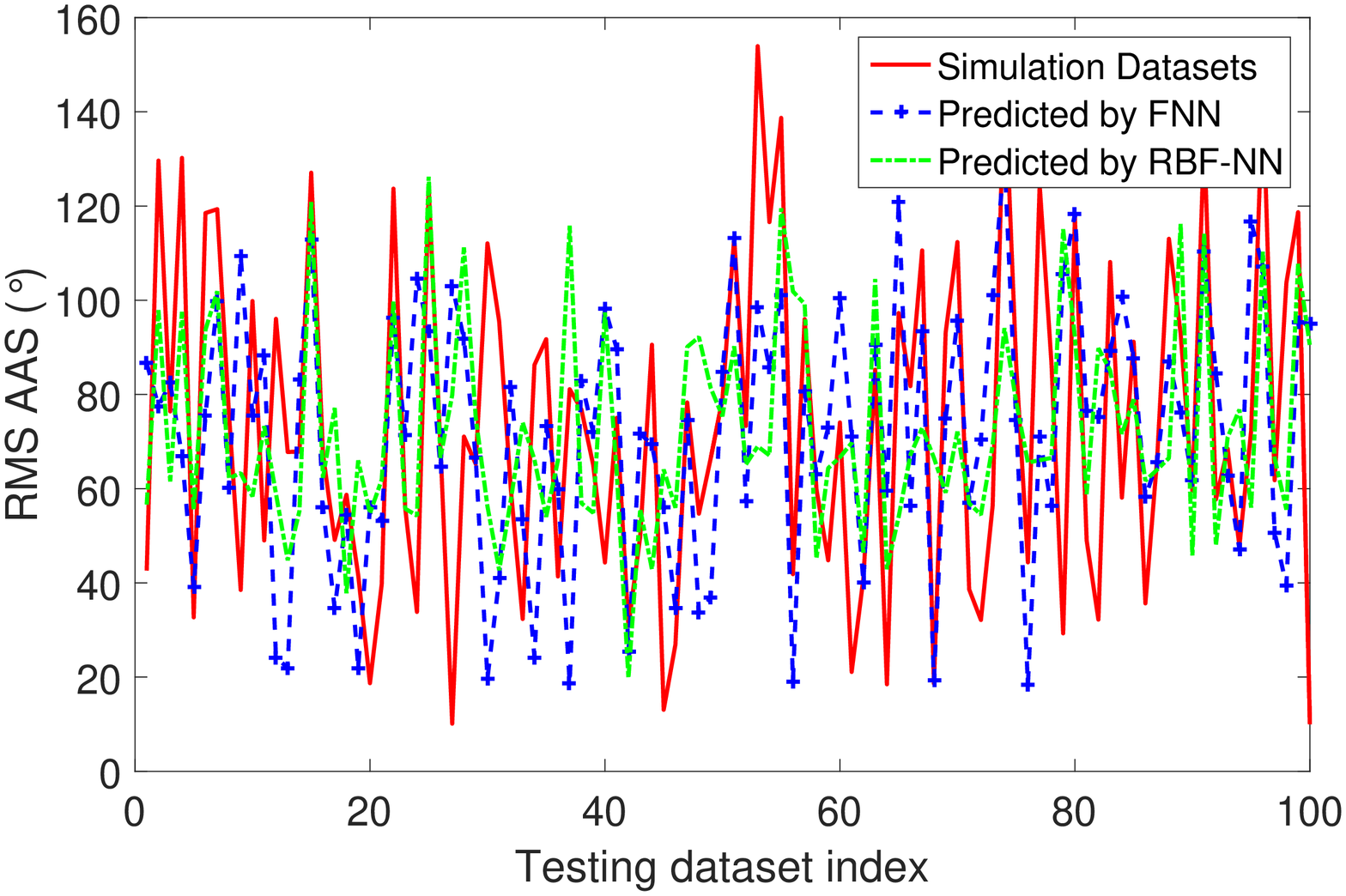}}
\footnotesize \centerline{(c)}
\end{minipage}
\begin{minipage}[t]{0.48\linewidth}
\centerline{\includegraphics[width=3in]{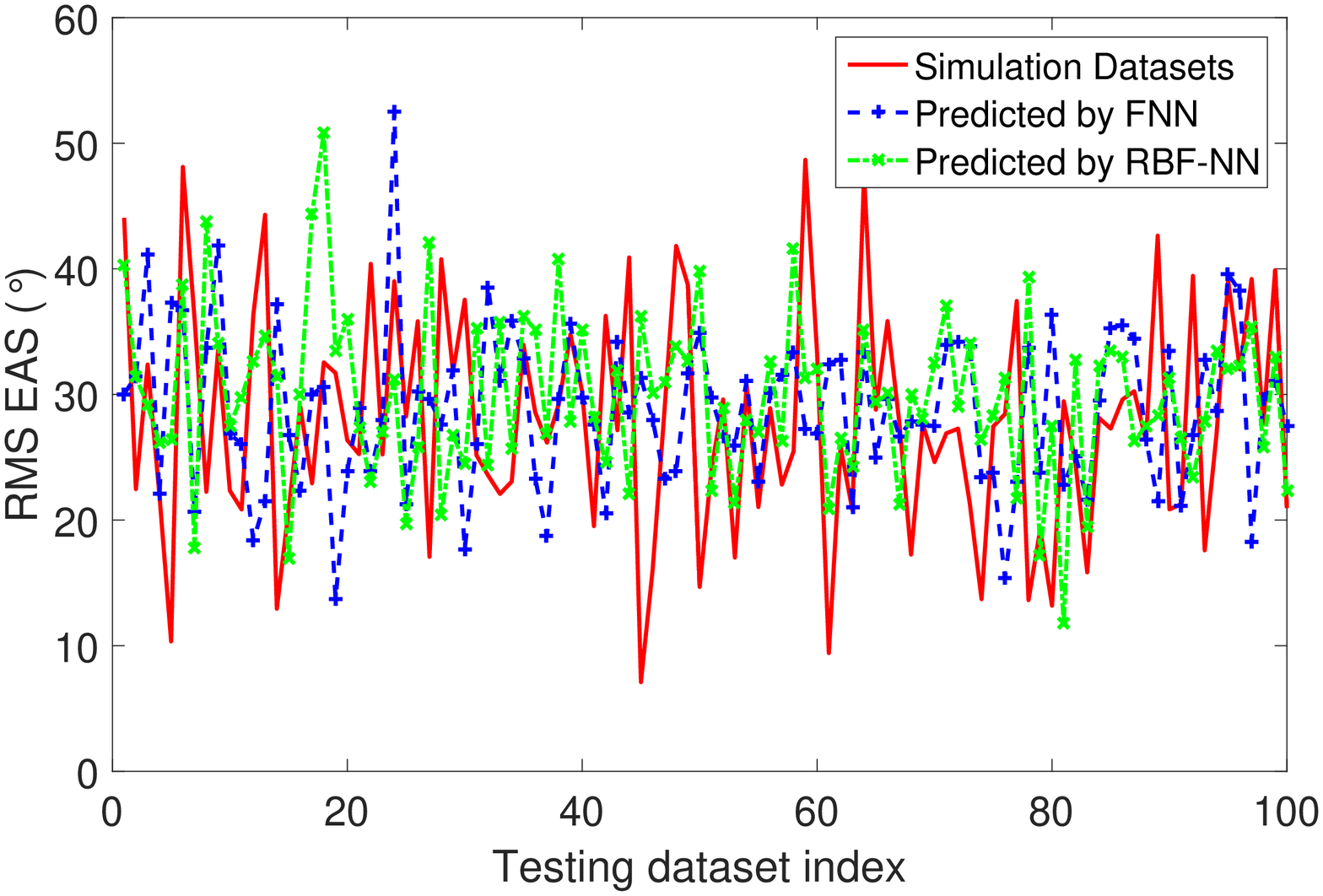}}
\footnotesize \centerline{(d)}
\end{minipage}
\caption{Simulated and predicted results of (a) RMS ADS, (b) RMS EDS, (c) RMS AAS, and (d) RMS EAS. }
\label{fig:sim_as}
\end{figure*}

In Fig. \ref{fig:sim_pl}, the simulated and predicted received powers are shown. Both the prediction performances of FNN and RBF-NN are illustrated. The x-axis denotes the index of testing dataset. The value of received power lies in the range of $-$80~dB to $-$60 dB. In Fig. \ref{fig:sim_ds}, the simulated and predicted RMS DSs are shown. The value of RMS DS lies in the range of 2~ns to 12 ns. The simulated and predicted RMS ASs including ADSs, EDSs, AASs, and EASs are shown in Fig.~\ref{fig:sim_as}. The azimuth angles are in the range of 20$^\circ$ to 140$^\circ$, while the elevation angles are in the range of 10$^\circ$ to 50$^\circ$. The GBSM simulation datasets have similar statistical properties with the channel measurement datasets.

In general, the value and trend of the output parameters can be predicted well, except that some points with too small or too large real values are predicted with relatively larger deviations. The prediction performances of received power and RMS DS are better than that of RMS ASs. The reason may be that the received power and RMS DS are more correlated with the input parameters than that of RMS ASs. Specifically, the received power is highly correlated with the Tx--Rx distance. Due to the fact that angle parameters of MPCs are more correlated with scatterers, the relationship between RMS AS and the input parameters may be weaker, thus leading to the worse performance. Moreover, as the azimuth ASs are varying in a larger range, their prediction performance is worse than that of elevation ASs.

In comparison, the outputs of the RBF-NN can have stable values, while a drawback of the FNN is that the outputs may be local optimization values. Thus, the outputs of the FNN can change for each run due to the initialization procedure of the neural network. Another problem for FNN is that the goal of convergence error may not be achieved. For those reasons, the performance of RBF-NN is generally better than that of FNN.

Compared with existing similar works \cite{Chang97, Azp14, Fer16_2}, the developed ANN based channel model framework is more robust and adaptive. Not only heights of Tx and Rx antennas are considered, but also their accurate locations in the horizontal plane. The carrier frequency are also varying in a large range, which covers most of the mmWave frequency bands like 11, 16, 28, 38, 45, and 60 GHz. Moreover, the output parameters are not limited to only the received power or the path loss. The channel sparsity properties in delay and 3-D double-directional angular domains are also considered.

\subsection{Prediction Performance Comparison}
Note that the goal of convergence error will have important impact on the performance of neural network. If the goal is near to zero, the learned neural network will be excellent for the training datasets. However, for the testing datasets, there may be over-fitting problems. That is to say, for some inputs with extreme values which are unknown to the learned neural network, the outputs may have great deviation from the expected values. Moreover, the predicted values may out of their reasonable ranges. Thus, in our simulations, the goal for FNN and RBF-NN are tuned to achieve good performance.

In addition, the root mean square errors (RMSEs) for FNN and RBF-NN are shown in Table \ref{tab:rmse}. RMSEs for the total dataset, training dataset, and testing dataset are given. As illustrated, the prediction of received power and RMS DS are better than that of RMS ASs. The performance of RBF-NN is better than that of FNN. Compared with RMSE for training datasets, the values of RMSE are a little larger for the total datasets and testing datasets.

\begin{table}[!bt]
\caption{RMSEs comparison for FNN and RBF-NN.}
\label{tab:rmse}
\begin{center}
\begin{tabular}{|p{2.4cm}|p{2.2cm}|p{2.2cm}|}
\hline
RMSE (total/training/testing)&FNN&RBF-NN\\
\hline
Received power (dB)&2.21/2.05/2.63 &1.98/1.46/3.05 \\
\hline
RMS DS (ns)&1.64/1.50/2.00 & 1.12/0.83/1.71\\
\hline
RMS ADS ($^\circ$)&24.94/21.28/33.61&19.92/15.05/30.13\\
\hline
RMS EDS ($^\circ$)& 9.88/8.46/13.24&7.71/5.11/12.63\\
\hline
RMS AAS ($^\circ$)&27.03/23.65/35.29&21.18/15.71/32.48\\
\hline
RMS EAS ($^\circ$)&9.38/8.67/11.26 &6.74/4.68/10.76  \\
\hline
\end{tabular}
\end{center}
\end{table}

\subsection{Input and Output Relationships}
\begin{figure}[tb!]
\centering
\includegraphics[width=3.4in]{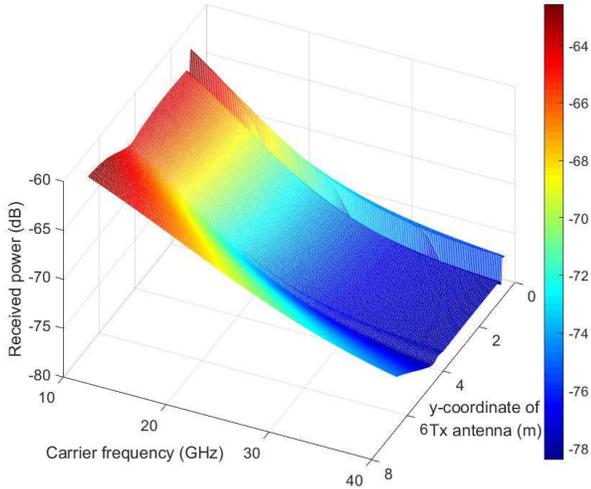}
\caption{The received power varying with carrier frequency and y-coordinate of Tx antenna predicted by FNN.}
\label{fig:power_fnn}
\end{figure}

\begin{figure}[tb!]
\centering
\includegraphics[width=3.4in]{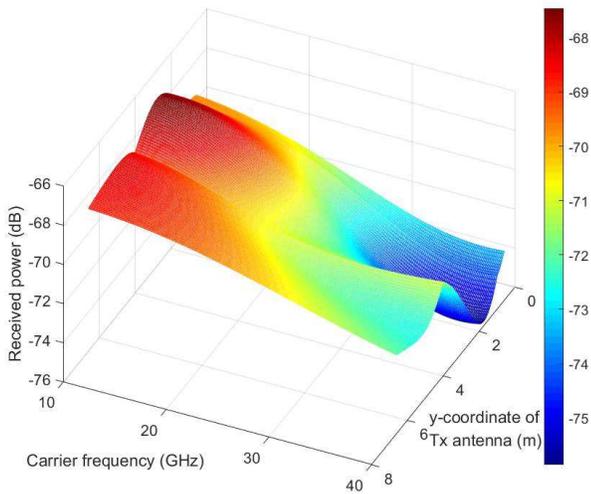}
\caption{The received power varying with carrier frequency and y-coordinate of Tx antenna predicted by RBF-NN.}
\label{fig:power_rbf}
\end{figure}

Once the neural network is trained and learned well, the non-linear relationships between inputs and outputs can be obtained. The six output parameters are distributed in a eight-dimensional space constructed by the eight input parameters. Here a simplified configuration is shown. The coordinate of Tx antenna is set as (1,3,1.45). The x-coordinate and z-coordinate of Rx antenna are set as 4 m and 2.6 m, respectively. The y-coordinate of Rx antenna is varying from 0.6 m to 6.6 m, and the Tx--Rx distance is also varying with their coordinates. The carrier frequency is varying from 10~GHz to 40 GHz. As an example, the influence of carrier frequency and y-coordinate of Tx antenna on the received power is analyzed. The received powers predicted by FNN and RBF-NN are shown in Fig. \ref{fig:power_fnn} and Fig. \ref{fig:power_rbf}, respectively. Generally, the FNN and RBF-NN predicted received powers show similar trend with varying input parameters. The curve surfaces are more smooth for RBF-NN. As the carrier frequency increases, the received power tends to be smaller, which is reasonable. As the y-coordinate of Tx antenna varies, the Tx--Rx distance also varies, the received power shows variations along the y-axis.

\subsection{Extensions and Discussions}
In this paper, the training and testing datasets are collected both from real channel measurement data and GBSM simulation data in an indoor environment. Both the FNN and RBF-NN are applied and validated to predict important channel statistical properties. In general, FNN and RBF-NN show similar performance on measurement datasets, while RBF-NN shows better performance than that of FNN for simulation datasets. Though the ANN based channel model framework is only applied to a specific indoor office environment, it can be extended to other more complicated scenarios. Datasets collected from different communication environments can be used together to train the ANN, thus obtaining a unified channel model framework. The learned channel model framework can also be used to classify different scenarios by distinguishing the pattern of channel statistical properties. Problems and challenges lie in three aspects:

\subsubsection{Scenario Representation}
The wireless channel is actually the physical medium in the environment. How to represent the influence of the environment is a challenging task. For 5G wireless communications, scenarios will be versatile, such as indoor hotspot, outdoor vehicle-to-vehicle (V2V), and high speed train (HST). For indoor scenarios, tables, desks, and electronic devices are some typical objects. For outdoor scenarios, trees, buildings, and cars are some typical objects instead. It is needed to find a general representation of the different objects in various scenarios to train big datasets obtained from different environments. Moreover, as different datasets from different scenarios will show different channel statistical properties or channel fingerprint, scenario classification can also be completed from learning of the big datasets.

\subsubsection{Big Datasets Acquisition}
We use both measured and simulated datasets in this paper. The real channel measurement data shows better performance. The influence of objects in the measured environment can be better reflected by real channel measurement data, and a more complicated relationship between input and output parameters may be learned. This indicates that channel sounder design and channel measurement campaigns will be very important for future 5G wireless channel modeling when applying machine learning based approaches. As channel measurements in various environments are expensive and time consuming, some attempts have been made to develop open access big channel measurement datasets to accelerate researches. For example, Shanghai Research Center for Wireless Communications (WiCO), which is a leading research and evaluation institute focusing on the R\&D of the key technologies for 5G wireless communication network, has conducted a large mount of wireless channel measurements in many typical scenarios, including beach, stadium, hotel lobby, rural area, campus hot spots, etc. More real measurement data of MIMO channels at different communication scenarios can be found at the open-source website \url{http://www.wise.sh/}.

\subsubsection{Inputs and Outputs Selection}
Actually, machine learning is to find the relationship between inputs and outputs. Thus, how to select the input and output parameters will have influence on the learned machine. As seen from our analysis, the received power and RMS DS have higher correlation with the Tx and Rx coordinates, Tx--Rx distance, and carrier frequency than that of RMS ASs. To achieve good performance, the most correlated parameters which may influence the outputs should be selected.

\section{Conclusion}
\label{sec_conclusion_section}
The exponential increase of mobile data has brought 5G wireless network to the era of big data. In this paper, we have investigated various machine learning algorithms and recent developments in applying big data analytics to wireless communications and channel modeling. An ANN based channel model framework has been proposed. Datasets have been collected from real channel measurements and GBSM simulations. Both the FNN and RBF-NN have been applied and compared. Important channel statistical properties including the received power, RMS DS, and RMS ASs have been predicted. Simulation results have been analyzed and validated, which have shown that machine learning algorithms can be powerful analytical tools for future measurement-based wireless channel modeling.

\section*{Acknowledgment}
\label{Thanks}
The authors would like to acknowledge the support from the National Key R\&D Program of China (Grant No. 2018YFB1801101 and YFB0102104),  the Natural Science Foundation of China (Grant No. 61960206006, 61901109, 61932014, 61771293, 61572323, and 61932014), the National Postdoctoral Program for Innovative Talents (Grant No. BX20180062), the Research Fund of National Mobile Communications Research Laboratory, Southeast University (Grant No. 2020B01), EU H2020 RISE TESTBED2 project (Grant No. 872172), the Fundamental Research Funds for the Central Universities (Grant No. 2242019R30001), the Huawei Cooperation Project, Science and Technology Commission of Shanghai Municipality (STCSM) (Grant No. 18511106500), and The Academy of Finland (Grant No. 319484).


%

\ifCLASSOPTIONcompsoc
\else
\fi

\ifCLASSOPTIONcaptionsoff
  \newpage
\fi

\begin{IEEEbiography}[{\includegraphics[width=1in,height=1.25in,clip,keepaspectratio]{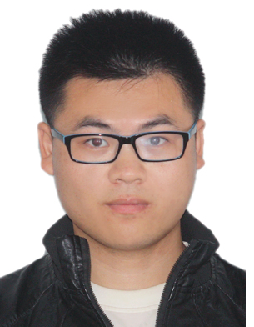}}]
{Jie Huang}received the B.E. degree in Information Engineering from Xidian University, China, in 2013, and the Ph.D. degree in Communication and Information Systems from Shandong University, China, in 2018. He is currently a Postdoctoral Research Associate in the National Mobile Communications Research Laboratory, Southeast University, China, and also with the Purple Mountain Laboratories, China. He received the Best Student Paper Award from the WPMC 2016. His research interests include millimeter wave and massive MIMO channel measurements, parameter estimation, channel modeling, wireless big data, and B5G/6G wireless communications.
\end{IEEEbiography}

\begin{IEEEbiography}[{\includegraphics[width=1in,height=1.25in,clip,keepaspectratio]{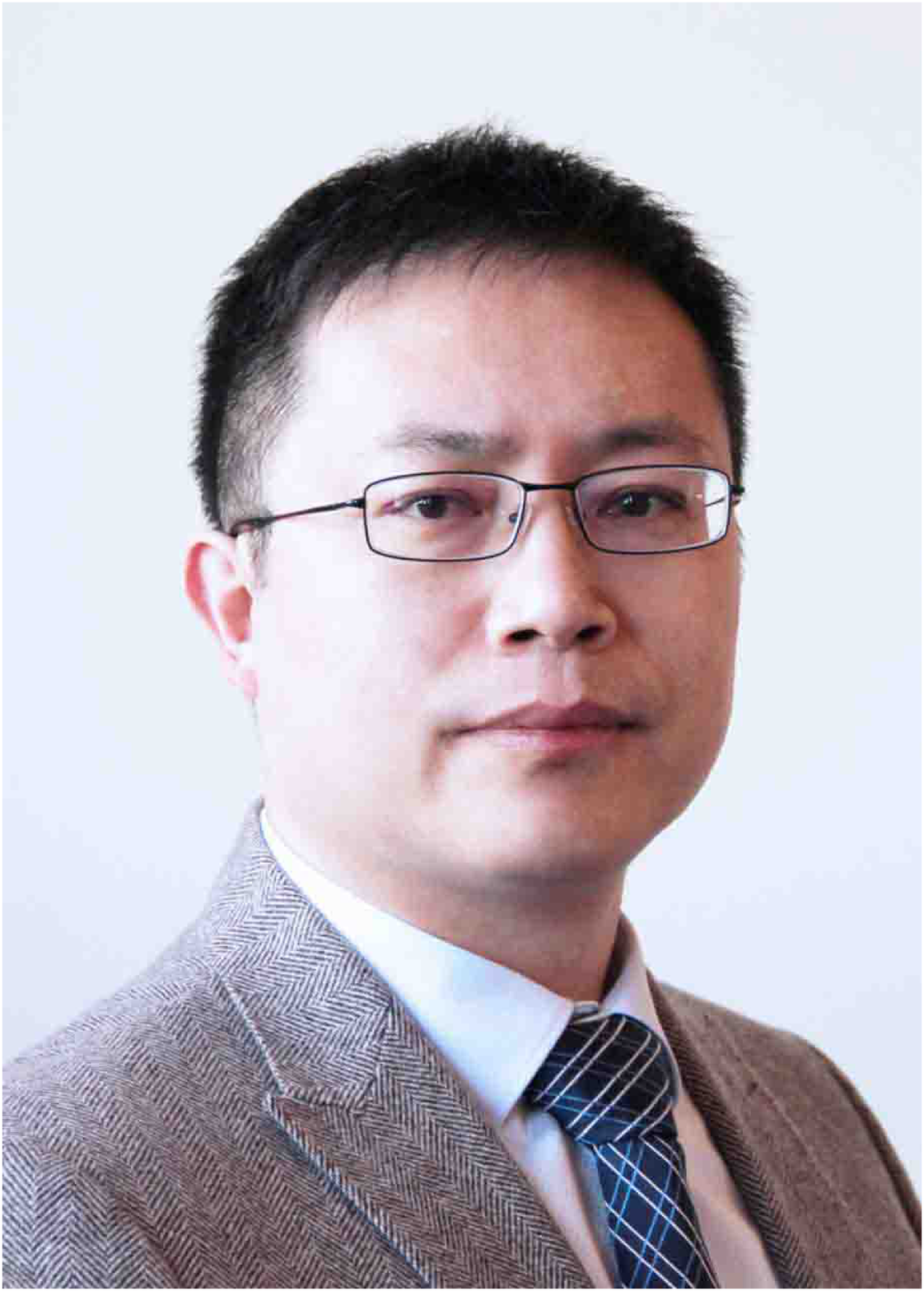}}]
{Cheng-Xiang Wang  (S'01-M'05-SM'08-F'17) }
received the BSc and MEng degrees in Communication and Information Systems from Shandong University, China, in 1997 and 2000, respectively, and the PhD degree in Wireless Communications from Aalborg University, Denmark, in 2004.

He was a Research Assistant with the Hamburg University of Technology, Hamburg, Germany, from 2000 to 2001, a Visiting Researcher with
Siemens AG Mobile Phones, Munich, Germany, in 2004, and a Research
Fellow with the University of Agder, Grimstad, Norway, from 2001 to 2005. He has been with Heriot-Watt University, Edinburgh, U.K., since 2005, where he was promoted to the position of Professor, in 2011. In 2018, he joined Southeast University, China, as a Professor. He is also a part-time professor with the Purple Mountain Laboratories, Nanjing, China. He has authored three books, one book chapter, and more than 370 articles in refereed journals and conference proceedings, including 23 Highly Cited Papers. He has also delivered 18 invited keynote speeches/talks and seven tutorials in international conferences. His current research interests include wireless channel measurements and modeling, B5G wireless communication networks, and applying artificial intelligence to wireless communication networks.

Dr. Wang is a Fellow of the IET, an IEEE Communications Society Distinguished Lecturer, in 2019 and 2020, and a Highly-Cited Researcher recognized by Clarivate Analytics, from 2017 to 2019. He is currently an Executive Editorial Committee Member of the IEEE TRANSACTIONS ON WIRELESS COMMUNICATIONS. He has served as an Editor for nine international journals, including the IEEE TRANSACTIONS ON WIRELESS COMMUNICATIONS, from 2007 to 2009, the IEEE TRANSACTIONS ON VEHICULAR TECHNOLOGY, from 2011 to 2017, and the IEEE TRANSACTIONS ON COMMUNICATIONS, from 2015 to 2017. He was a Guest Editor of the IEEE JOURNAL ON SELECTED AREAS IN COMMUNICATIONS, the Special Issue on Vehicular Communications and Networks (Lead Guest Editor), the Special Issue on Spectrum and Energy Efficient Design of Wireless Communication Networks, and the Special Issue on Airborne Communication Networks. He was also a Guest Editor of the IEEE TRANSACTIONS ON BIG DATA, the Special Issue on Wireless Big Data, and is a Guest Editor of the IEEE TRANSACTIONS ON COGNITIVE COMMUNICATIONS AND NETWORKING, the Special Issue on Intelligent Resource Management for 5G and Beyond. He has served as a TPC member, the TPC Chair, and the General Chair for more than 80 international conferences. He received ten Best Paper Awards from the IEEE GLOBECOM 2010, IEEE ICCT 2011, ITST 2012, IEEE VTC 2013-Spring, IWCMC 2015, IWCMC 2016, IEEE/CIC ICCC 2016, WPMC 2016, and WOCC 2019.
\end{IEEEbiography}

\begin{IEEEbiography}[{\includegraphics[width=1in,height=1.25in,clip,keepaspectratio]{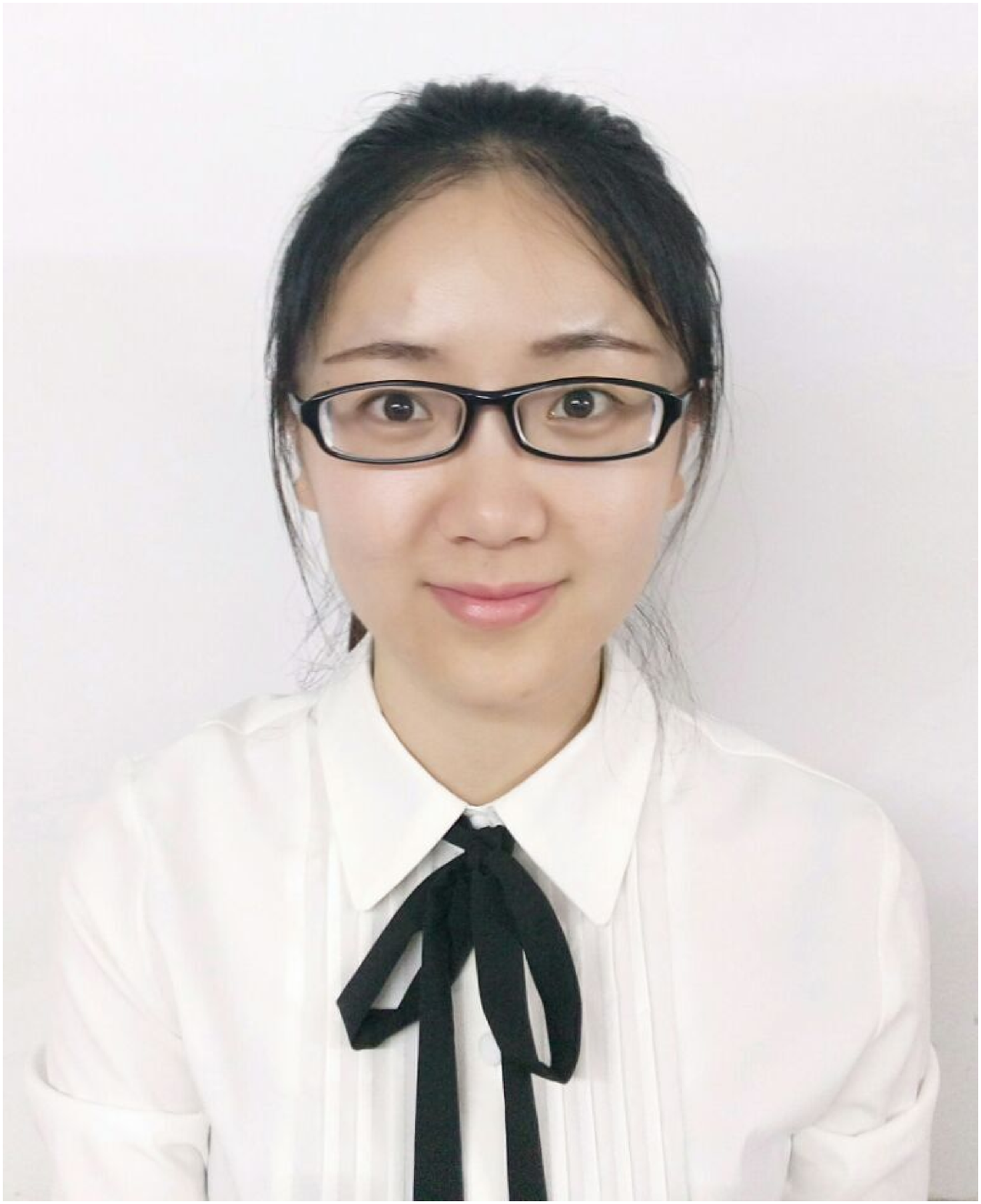}}]
{Lu Bai}received the B.Sc. degree in Electronic Information Engineering from Qufu Normal University, China, in 2014, and the Ph.D. degree in Information and Communication Engineering from Shandong University, China, in 2019. From 2017 to 2019, she was also a visiting Ph.D. student at Heriot-Watt University, UK. Now, she is a research fellow at Beihang University, China. Her research interests include massive MIMO channel measurements and modeling, satellite communication channel modeling, and wireless big data. 
\end{IEEEbiography}

\begin{IEEEbiography}[{\includegraphics[width=1in,height=1.25in,clip,keepaspectratio]{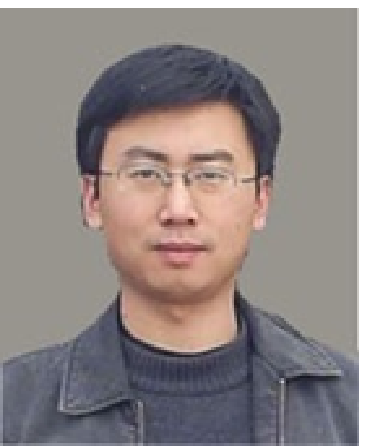}}]
{Jian Sun (M'08)}received the PhD degree from Zhejiang University, Hangzhou, China, in March 2005.
Since July 2005, he has been a Lecturer in the School of Information Science and Engineering, Shandong University, China. In 2011, he was a visiting scholar at Heriot-Watt University, UK, supported by UK-China Science Bridges: R\&D on (B)4G Wireless Mobile Communications (UC4G) project.
His research interests are in the areas of signal processing for wireless communications, channel sounding and modeling, propagation measurement and parameter extraction, MIMO and multicarrier transmission systems design and implementation.
\end{IEEEbiography}

\begin{IEEEbiography}[{\includegraphics[width=1in,height=1.25in,clip,keepaspectratio]{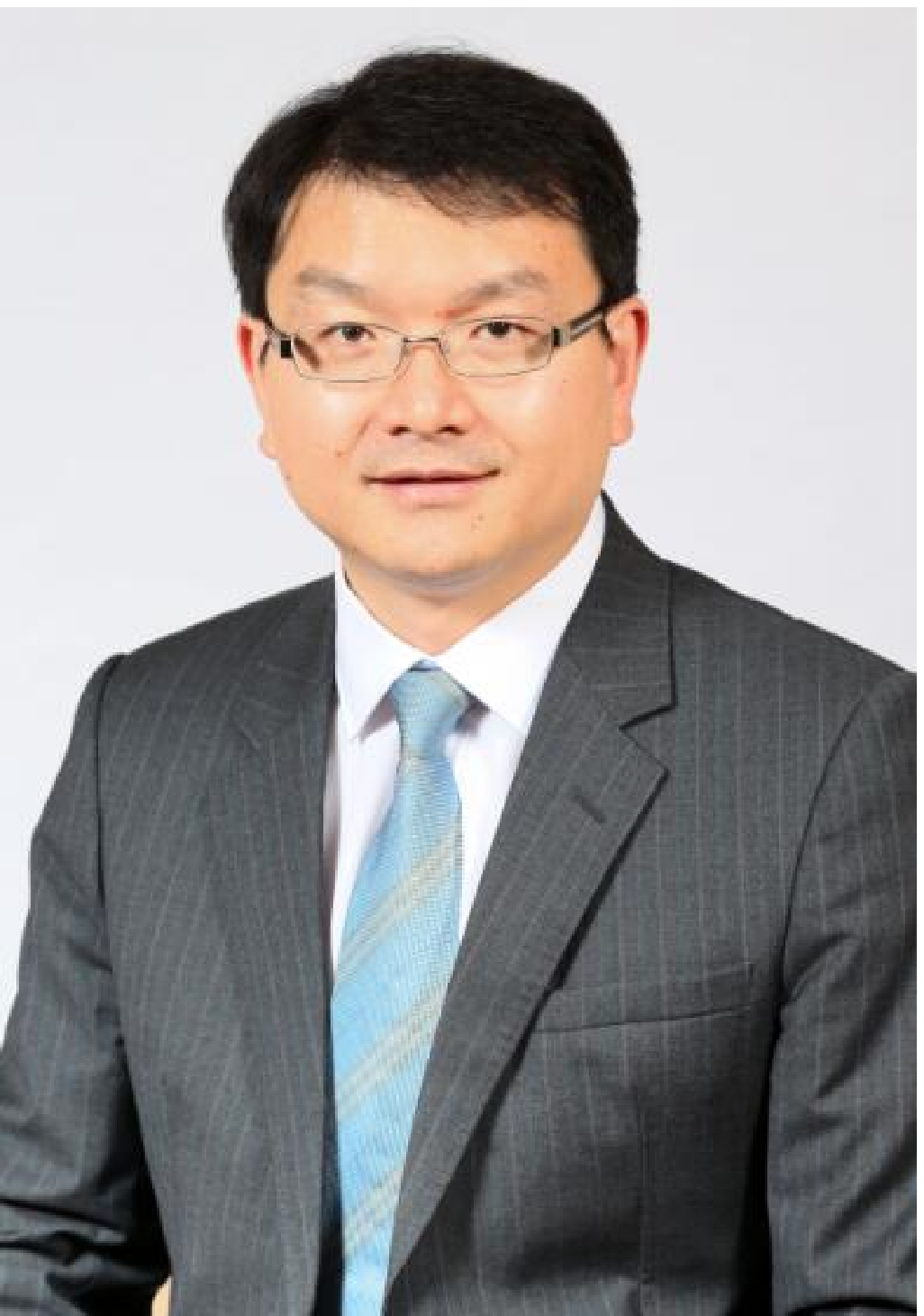}}]
{Yang Yang (S'99-M'02-SM'10-F'18)}
received the BEng and MEng degrees in Radio Engineering from Southeast University, Nanjing, China, in 1996 and 1999, respectively, and the PhD degree in Information Engineering from The Chinese University of Hong Kong in 2002.

Dr. Yang Yang is currently a full professor with School of Information Science and Technology, ShanghaiTech University, China, serving as the Executive Dean of School of Creativity and Art, and the Co-Director of
Shanghai Institute of Fog Computing Technology (SHIFT). Prior to that, he has held faculty positions at The Chinese University of Hong Kong,
Brunel University (UK), University College London (UCL, UK), and Shanghai Institute of Microsystem and Information Technology (SIMIT), Chinese Academy of Sciences (CAS, China). He is a member of the Chief Technical Committee of the National Science and Technology Major Project ``New Generation Mobile Wireless Broadband Communication Networks'' (2008-2020), which is funded by the Ministry of Industry and Information Technology (MIIT) of China. In addition, he is on the Chief Technical Committee for the National 863 Hi-Tech R\&D Program ``5G System R\&D Major Projects'', which is funded by the Ministry of Science and Technology (MOST) of China. His current research interests include fog computing networks, service-oriented collaborative intelligence, wireless sensor networks, IoT applications, and advanced testbeds and experiments. He has published more than 200 papers and filed more than 80 technical patents in these research areas. He is the Chair of the Steering Committee of Asia-Pacific Conference on Communications (APCC) since January 2019. Yang is a Fellow of the IEEE. He is a General Co-Chair of IEEE DSP 2018 conference and a TPC Vice-Chair of IEEE ICC 2019 conference.
\end{IEEEbiography}

\begin{IEEEbiography}[{\includegraphics[width=1in,height=1.25in,clip,keepaspectratio]{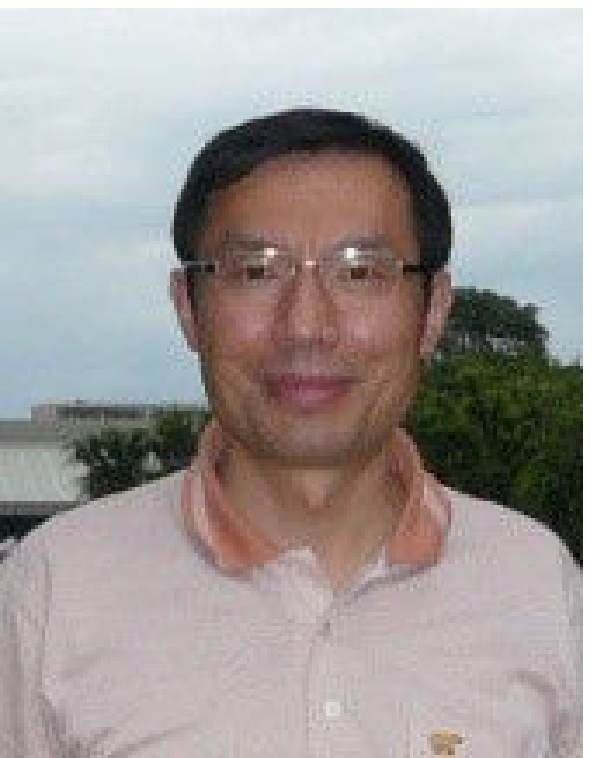}}]
{Jie Li }
received the BSc degree in Computer Science from Zhejiang University, Hangzhou, China, the MEng degree in Electronic Engineering and Communication systems from China Academy of Posts and Telecommunications, Beijing, China. He received the PhD degree from the University of Electro-Communications, Tokyo, Japan. He is with Faculty of Engineering, Information and Systems, University
of Tsukuba, Japan, where he is a Professor. He has been a visiting professor in Yale University, USA, Inria France. His current research interests are in mobile distributed computing and networking, big data and cloud computing, IoT, OS, modeling and performance evaluation of information systems. He is a senior member of IEEE and ACM and a member of IPSJ (Information
Processing Society of Japan).
\end{IEEEbiography}

\begin{IEEEbiography}[{\includegraphics[width=1in,height=1.25in,clip,keepaspectratio]{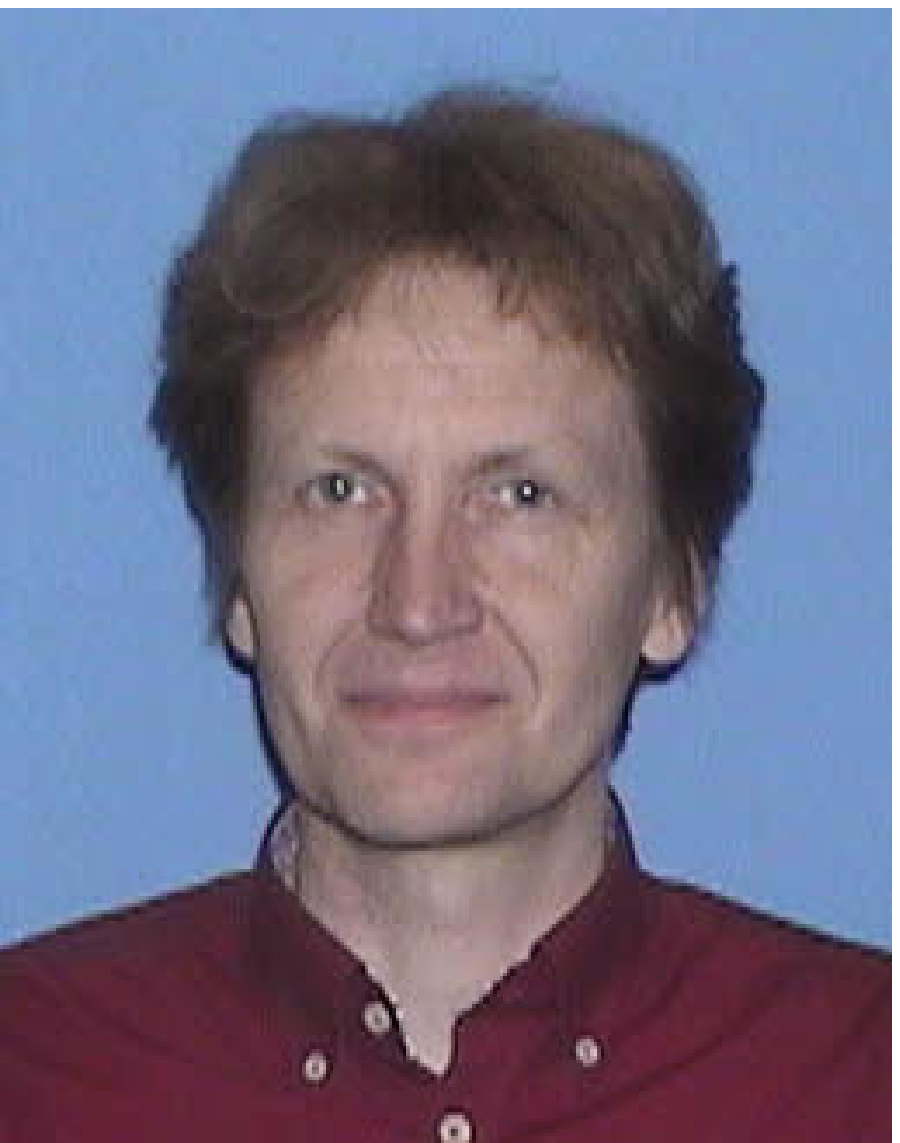}}]
{Olav Tirkkonen }
received the MEng and PhD degrees in Theoretical Physics from Helsinki University of Technology, Espoo, Finland, in 1990 and 1994, respectively. He is currently an Associate Professor of
communication theory with the Department of Communications and Networking, Aalto University, Espoo, where he has held a faculty position since August 2006. Between 1994 and 1999, he held Postdoctoral positions with the University of British Columbia,
Vancouver, BC, Canada, and the Nordic Institute for Theoretical Physics, Copenhagen, Denmark. From 1999 to 2010, he was with Nokia Research Center, Helsinki, Finland. He has published more than 200 papers and is coauthor of the book Multiantenna Transceiver Techniques for 3G and Beyond (Wiley, 2004). His current research
interests include coding theory, multiantenna techniques, and cognitive management of fifth-generation cellular networks.
\end{IEEEbiography}

\begin{IEEEbiography}[{\includegraphics[width=1in,height=1.25in,clip,keepaspectratio]{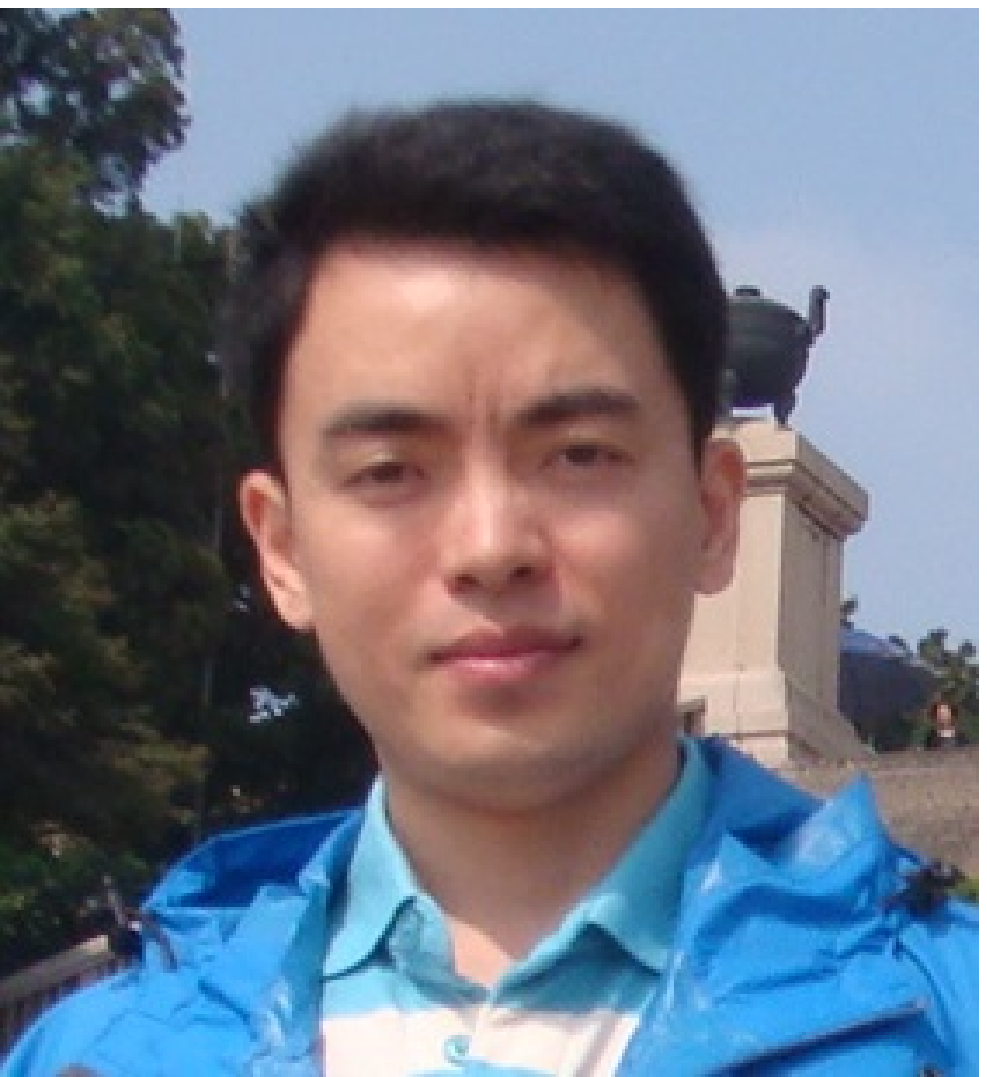}}]
{Ming-Tuo Zhou ((S'01–M'04–SM'11))}
received the BSc degree from Hunan University in 1997, MEng degree from Chongqing University of Posts and Telecommunications in 2000, and PhD degree from Asian Institute of Technology in 2003. He joined Shanghai Institute of Microsystem and Information Technology (SIMIT), Chinese Academy of Sciences (CAS) in September 2016, and now is the head of the Computing \& Communication Group and a professor. He is also affiliated with the Shanghai Institute of Fog Computing Technology (SHIFT), ShanghaiTech University, China. He was a senior research scientist at the Smart Wireless Laboratory of the (Japan) National Institute of Information and Communications Technology (NICT) Singapore Representative Office during July 2004 to August 2016. He was the Technical Co-Editor of IEEE 802.16n and IEEE 802.16.1a, a voting member and technical contributor of the IEEE 802.16, 802.15, 802.11, and 802.19 Working Groups. He was a member of the Test and Certification Working Group of Wi-SUN Alliance. He has co-authored about 80 technical papers, 8 book chapters, and co-edited two technical books. He has served as Technical Program Committee member, Finance Chair, Local Arrangement Chair, and Session Chair at more than 40 international conferences including ICC, GLOBECOM, and PIMRC. His current research interests include fog computing and communications, Internet of things, software-defined networks, and cyber-physical systems. He is IEEE SA member and IEEE Senior Member.
\end{IEEEbiography}


\begin{thebibliography}{1}

\bibitem{Cisco17} ``Cisco white paper,'' [Online]. Available:
\url{http://www.cisco.com/c/en/us/solutions/collateral/service-provider/visual-networking-index-vni/mobile-white-paper-c11-520862.pdf} [Accessed: 10-Mar-2017]

\bibitem{Wang14} C.-X. Wang, F. Haider, X. Gao, X.-H. You, Y. Yang, D. Yuan, H. Aggoune, H. Haas, S. Fletcher, and E. Hepsaydir, ``Cellular architecture and key technologies for 5G wireless communication networks,'' \emph{IEEE Commun. Mag.}, vol. 52, no. 2, pp. 122--130, Feb. 2014.

\bibitem{Yang17} Y. Yang, J. Xu, G. Shi, and C.-X. Wang, \emph{5G Wireless Systems: Simulation and Evaluation Techniques}. Chippenham: Springer, Oct. 2017.
 
\bibitem{And14} J. G. Andrews, S. Buzzi, W. Choi, S. V. Hanly, A. Lozano, A. C. K. Soong, and J. Zhang, ``What will 5G be?'' \emph{IEEE J. Sel. Areas Commun.}, vol. 32, no. 6, pp. 1065--1082, Jun. 2014.

\bibitem{ITU} ``ITU-R M.2083-0,'' [Online]. Available:\url{https://www.itu.int/dms_pubrec/itu-r/rec/m/R-REC-M.
2083-0-201509-I!!PDF-E.pdf} [Accessed: 10-Sept-2017]

\bibitem{Shafi17} M. Shafi, A. F. Molisch, P. J. Smith, T. Haustein, P. Zhu, P. De Silva, F. Tufvesson, A. Benjebbour, and G. Wunder, ``5G: A tutorial overview of standards, trials, challenges, deployment, and practice,'' \emph{IEEE J. Sel. Areas Commun.}, vol. 35, no. 6, pp. 1201--1221, Jun. 2017.

\bibitem{Ge16} X. Ge, S. Tu, G. Mao, C.-X. Wang, and T. Han, ``5G ultra-dense cellular networks,'' \emph{IEEE Wireless Commun.}, vol. 23, no. 1, pp. 72--79, Feb. 2016.

\bibitem{Bi15} S. Bi, R. Zhang, Z. Ding, and S. Cui, ``Wireless communications in the era of big data,'' \emph{IEEE Commun. Mag.}, vol. 53, no.~10, pp. 190--199, Oct.~2015.

\bibitem{Fer16} P. Ferrand, M. Amara, S. Valentin, and M. Guillaud, ``Trends and challenges in wireless channel modeling for evolving radio access,'' \emph{IEEE Commun. Mag.}, vol. 54, no. 7, pp. 93--99, Jul. 2016.

\bibitem{Gan15} A. Gandomi and M. Haider, ``Beyond the hype: Big data concepts, methods, and analytics,'' \emph{Int. J. Inf. Manage.}, vol. 35, no. 2, pp. 137--~144, Apr. 2015.

\bibitem{Katal13} A. Katal, M. Wazid, and R. H. Goudar, ``Big data: Issues, challenges, tools and good practices,'' in \emph{Proc. IC3'13}, Noida, India, Aug. 2013, pp. 404--409.

\bibitem{Chin14} W. H. Chin, Z. Fan, and R. Haines, ``Emerging technologies and research challenges for 5G wireless networks,'' \emph{IEEE Wireless Commun.}, vol. 21, no. 2, pp. 106--112, May 2014.

\bibitem{Cheng17} X. Cheng, L. Fang, X. Hong, and L. Yang, ``Exploiting mobile big data: Sources, features, and applications,'' \emph{IEEE Netw.}, vol. 31, no. 1, pp. 72--~79, Jan. 2017.

\bibitem{Cheng17_2} X. Cheng, L. Fang, and L. Yang, ``Mobile big data: The fuel for data-driven wireless,'' \emph{IEEE Int. Things}, vol. 4, no. 5, pp. 1489--~1516, Jun. 2017.

\bibitem{Imran14} A. Imran, A. Zoha, and A. Abu-Dayya, ``Challenges in 5G: How to empower SON with big data for enabling 5G,'' \emph{IEEE Netw.}, vol. 28, no.~6, pp. 27--33, Nov. 2014.

\bibitem{Han15} Q. Han, S. Liang, and H. Zhang, ``Mobile cloud sensing, big data, and 5G networks make an intelligent and smart world,'' \emph{IEEE Netw.}, vol. 29, no. 2, pp. 40--45, Mar. 2015.

\bibitem{Zheng16} K. Zheng, Z. Yang, K. Zhang, P. Chatzimisios, K. Yang, and W. Xiang, ``Big data-driven optimization for mobile networks toward 5G,'' \emph{IEEE Netw.}, vol. 30, no. 1, pp. 44--51, Jan. 2016.

\bibitem{Jiang16} C. Jiang, H. Zhang, Y. Ren, Z. Han, K.-C. Chen, and L. Hanzo, ``Machine learning paradigms for next-generation wireless networks,'' \emph{IEEE Wireless Commun.}, vol. 24, no. 2, pp. 98--105, Apr. 2017.

\bibitem{Han17} S. Han, C.-L. I, G. Li, S. Wang, and Q. Sun, ``Big data enabled mobile network design for 5G and beyond,'' \emph{IEEE Commun. Mag.}, vol. 55, no. 9, pp. 150--157, Sept. 2017.

\bibitem{Qian17} L. Qian, J. Zhu, and S. Zhang, ``Survey of wireless big data,'' \emph{J. Commun. Inf. Netw.}, vol. 2, no. 1, pp. 1--18, Mar. 2017.

\bibitem{Ahm18} E. Ahmed, I. Yaqoob, I. A. T. Hashem, J. Shuja, M. Imran, N. Guizani, and S. T. Bakhsh, ``Recent advances and challenges in mobile big data,'' \emph{IEEE Commun. Mag.}, vol. 56, no. 2, pp. 102--108, Feb. 2018.

\bibitem{Zhang18} N. Zhang, P. Yang, J. Ren, D. Chen, L. Yu and X. Shen, ``Synergy of big data and 5G wireless networks: Opportunities, approaches, and challenges,'' \emph{IEEE Wireless Commun.}, vol. 25, no. 1, pp. 12--18, Feb. 2018.

\bibitem{Mar10} S. Maran\`o, W. M. Gifford, H. Wymeersch, and M. Z. Win, ``NLOS identification and mitigation for localization based on UWB experimental data,'' \emph{IEEE J. Sel. Areas Commun.}, vol. 28, no. 7, pp. 1026--1035, Sept.~2010.

\bibitem{Ngu15} T. V. Nguyen, Y. Jeong, H. Shin, and M. Z. Win, ``Machine learning for wideband localization,'' \emph{IEEE J. Sel. Areas Commun.}, vol. 33, no. 7, pp.~1357--1380, Jul. 2015.

\bibitem{Zou16} H. Zou, B. Huang, X. Lu, H. Jiang, and L. Xie, ``A robust indoor positioning system based on the procrustes analysis and weighted
extreme learning machine,'' \emph{IEEE Trans. Wireless Commun.}, vol. 15, no. 2, pp.~1252--1266, Feb. 2016.

\bibitem{Liang17} X. Liang, H. Zhang, T. Lu, and T. A. Gulliver, ``Extreme learning machine for 60 GHz millimetre wave positioning,'' \emph{IET Commun.}, vol.~11, no. 4, pp. 483--489, Mar. 2017.

\bibitem{Ye17} X. Ye, X. Yin, X. Cai, A. P. Yuste, and H. Xu, ``Neural-network-assisted UE localization using radio-channel fingerprints in LTE networks,'' \emph{IEEE Access}, vol.~5, pp. 12071--12087, Jun. 2017.

\bibitem{Thi13} K. M. Thilina, K. W. Choi, N. Saquib, and E. Hossain, ``Machine learning techniques for cooperative spectrum sensing in cognitive radio networks,'' \emph{IEEE J. Sel. Areas Commun.}, vol. 31, no. 11, pp. 2209--2221, Nov. 2013.

\bibitem{Joung16} J. Joung, ``Machine learning-based antenna selection in wireless communications,'' \emph{IEEE Commun. Lett.}, vol. 20, no. 11, pp. 2241--2244, Nov.~2016.

\bibitem{He18} D. He, C. Liu, T. Q. S. Quek, and H. Wang, ``Transmit antenna selection in MIMO wiretap channels: A machine learning approach,'' \emph{IEEE Wireless Commun. Lett.}, vol. 7, no. 4, pp. 634--637, Aug.~2018.



\bibitem{Bas14} E. Bastu\v g, M. Bennis, and M. Debbah, ``Living on the edge:
The role of proactive caching in 5G wireless networks,'' \emph{IEEE Commun. Mag.}, vol.~52, no. 8, pp. 82--89, Aug. 2014.

\bibitem{Bas15} E. Bastu\v g, M. Bennis, E. Zeydan, M. A. Kader, I. A. Karatepe, A. S. Er, and M. Debbah, ``Big data meets telcos: A proactive caching perspective,'' \emph{J. Commun. Netw.}, vol. 17, no. 6, pp. 549--557, Dec. 2015.

\bibitem{Yao16} H. Yao, C. Qiu, C. Fang, X. Chen, and F. R. Yu, ``A novel framework of data-driven networking,'' \emph{IEEE Access}, vol. 4, pp. 9066--9072, Nov.~2016.

\bibitem{Yang16} K. Yang, Q. Yu, S. Leng, B. Fan, and F. Wu, ``Data and energy integrated communication networks for wireless big data,'' \emph{IEEE Access}, vol. 4, pp.~713--723, Feb. 2016.

\bibitem{Huang17} Y. Huang, J. Tan, and Y.-C. Liang, ``Wireless big data: Transforming heterogeneous networks to smart networks,'' \emph{J. Commun. Inf. Netw.}, vol.~2, no. 1, pp. 19--32, Mar. 2017.

\bibitem{Bao17} Y. Bao, H. Wu, and X. Liu, ``From prediction to action: Improving user experience with data-driven resource allocation,'' \emph{IEEE J. Sel. Areas Commun.}, vol. 35, no. 5, pp. 1062--1075, May 2017.

\bibitem{Deb15} S. Deb and P. Monogioudis, ``Learning-based uplink interference management in 4G LTE cellular systems,'' \emph{IEEE Trans. Netw.}, vol. 23, no. 2, pp. 398--411, Apr. 2015.

\bibitem{He18_2} H. He, C.-K. Wen, S. Jin, and G. Y. Li, ``Deep learning-based channel estimation for beamspace mmWave massive MIMO systems,'' \emph{IEEE Wireless Commun. Lett.}, vol. 7, no. 5, pp. 852--855, Oct. 2018.

\bibitem{Tang18} H. Tang, J. Wang, and L. He, ``Off-grid sparse Bayesian learning based channel estimation for mmWave massive MIMO uplink,'' \emph{IEEE Wireless Commun. Lett.}, vol. 8, no. 1, pp. 45--48, Feb. 2019.

\bibitem{Luo18} C. Luo, J. Ji, Q. Wang, X. Chen, and P. Li, ``Channel state information prediction for 5G wireless communications: A deep learning approach,'' \emph{IEEE Trans. Netw. Sci. Eng.}, 2018, in press.

\bibitem{Meng18} F. Meng, P. Chen, L. Wu, and X. Wang, ``Automatic modulation classification: A deep learning enabled approach,'' \emph{IEEE Trans. Veh. Technol.}, vol. 67, no. 11, pp. 10760--10772, Nov. 2018.

\bibitem{Alh18} M. I. AlHajri, N. T. Ali, and R. M. Shubair, ``Classification of indoor environments for IoT applications: A machine learning approach,'' \emph{IEEE Antennas Wireless Propag. Lett.}, vol. 17, no. 12, pp. 2164--2168, Dec. 2018.

\bibitem{Cui18} J. Cui, Z. Ding, P. Fan, and N. Al-Dhahir, ``Unsupervised machine learning based user clustering in millimeter-wave-NOMA systems,'' \emph{IEEE Trans. Wireless Commun.}, vol. 17, no. 11, pp. 7425--7440, Nov. 2018.

\bibitem{Chang97} P.-R. Chang and W.-H. Yang, ``Environment-adaptation mobile radio propagation prediction using radial basis function neural networks,'' \emph{IEEE Trans. Veh. Technol.}, vol. 46, no. 1, pp. 155--160, Feb. 1997.

\bibitem{Liu12} Z. Liu, Z. Huang, and Y. Zhou, ``An efficient maximum likelihood method for direction-of-arrival estimation via sparse Bayesian learning,'' \emph{IEEE Trans. Wireless Commun.}, vol. 11, no. 10, pp. 3607--3617, Oct.~2012.

\bibitem{Braz13} J. A. Cal-Braz, L. J. Matos, and E. Cataldo, ``The relevance vector machine applied to the modeling of wireless channels,'' \emph{IEEE Trans. Antennas Propag.}, vol. 61, no. 12, pp. 6157--6167, Dec. 2013.

\bibitem{Azp14} L. Azpilicueta, M. Rawat, K. Rawat, F. M. Ghannouchi, and F. Falcone, ``A ray launching-neural network approach for radio wave propagation analysis in complex indoor environments,'' \emph{IEEE Trans. Antennas Propag.}, vol. 62, no. 5, pp. 2777--2786, May 2014.


\bibitem{Zhang16} J. Zhang, ``The interdisciplinary research of big data and wireless channel: A cluster-nuclei based channel model,'' \emph{China Commun.}, vol.~13, no. 2, pp. 14--26, Dec. 2016.

\bibitem{Ma17} X. Ma, J. Zhang, Y. Zhang, and Z. Ma, ``Data scheme-based wireless channel modeling method: Motivation, principle and performance,'' \emph{J. Commun. Inf. Netw.}, vol. 2, no. 3, pp. 41--51, Sept. 2017.

\bibitem{Ma17_2} X. Ma, J. Zhang, Y. Zhang, Z. Ma, and Y. Zhang, ``A PCA-based modeling method for wireless MIMO channel,'' in \emph{Proc. IEEE INFOCOM Wkshps'17}, Atlanta, GA, USA, May 2017, pp. 874--879.

\bibitem{Fer16_2} G. P. Ferreira, L. J. Matos, and J. M. M. Silva, ``Improvement of outdoor signal strength prediction in UHF band by artificial neural network,'' \emph{IEEE Trans. Antennas Propag.}, vol. 64, no. 12, pp. 5404--5410, Dec. 2016.

\bibitem{Li17} H. Li, Y. Li, S. Zhou, and J. Wang, ``Wireless channel feature extraction via GMM and CNN in the tomographic channel model,'' \emph{J. Commun. Inf. Netw.}, vol. 2, no. 1, pp. 41--51, Mar. 2017.

\bibitem{Bla17} T. Blazek and C. F. Mecklenbr\"auker, ``Sparse time-variant impulse response estimation for vehicular channels using the c-LASSO,'' in \emph{Proc. IEEE PIMRC'17}, Montreal, Canada, Oct. 2017, pp. 1--5.

\bibitem{Ucc18} M. Uccellari, F. Facchini, M. Sola, E. Sirignano, G. M. Vitetta, A. Barbieri, and S. Tondelli, ``On the application of support vector machines to the prediction of propagation losses at 169 MHz for smart metering applications,'' \emph{IET Microw. Antennas Propag.}, vol. 12, no. 3, pp. 302--312, Feb. 2018. 

\bibitem{He18_3} R. He, B. Ai, A. F. Molisch, G. L. St\"uber, Q. Li, Z. Zhong, and J. Yu, ``Clustering enabled wireless channel modeling using big data algorithms,'' \emph{IEEE Commun. Mag.}, vol. 56, no. 5, pp. 177--183, May 2018.


\bibitem{JSAC17} J. Huang, C.-X. Wang, R. Feng, J. Sun, W. Zhang, and Y. Yang, ``Multi-frequency mmWave massive MIMO channel measurements and characterization for 5G wireless communication systems,'' \emph{IEEE J. Sel. Areas Commun.}, vol. 35, no. 7, pp. 1591--1605, Jul. 2017.

\bibitem{SCIS18} J. Huang, C.-X. Wang, Y. Liu, J. Sun, and W. Zhang, ``A novel 3D GBSM for mmWave MIMO channels,'' \emph{Sci. China Inf. Sci.}, vol. 61, no. 102305, pp. 1--15, Oct. 2018.

\end{thebibliography}
\end{document}